\documentclass[12pt,letterpaper]{article}
\usepackage{jheppub}
\usepackage{hyperref}
\usepackage{empheq}
\RequirePackage{pgfopts}[2011/06/02]
\usepackage{amssymb}
\usepackage{color}
\usepackage{slashed}
\usepackage{amsmath}
\usepackage{amsfonts}
\usepackage{amssymb}
\usepackage{graphicx,subcaption}
\usepackage{caption}
\usepackage{bbold}
\usepackage{subcaption}
\usepackage{float}
\usepackage{tikz}
\usetikzlibrary[arrows]
\usepackage{amsmath}
\usepackage{epsfig}
\usepackage{appendix}
\usepackage{listings}
\usepackage{xcolor}
\usepackage{empheq}
\usepackage{ytableau}

\ytableausetup{boxsize=5pt}

\newcommand{\vev}[1]{{\left< {#1} \right>}}
\newcommand{\be}{\begin{equation}}
\newcommand{\ee}{\end{equation}}

\def\bZ {\mathbb{Z}}

\newcommand{\gYM}{g_{\scriptscriptstyle{\mathrm{YM}}}}

\title{Wilson loops in terms of color invariants}

\author{Bartomeu Fiol,}
\author{Jairo Mart\'inez-Montoya}
\author{and Alan Rios Fukelman}

\affiliation{Departament de F{\'\i}sica Qu\`antica i Astrof\'isica i \\Institut de Ci{\`e}ncies del Cosmos, 
Universitat de Barcelona,
Mart{\'\i}\ i Franqu{\`e}s 1, 08028 Barcelona, Catalonia, Spain}

\emailAdd{bfiol@ub.edu}
\emailAdd{jmartinez@icc.ub.edu}
\emailAdd{ariosfukelman@icc.ub.edu}

\abstract{We derive an expression for the vacuum expectation value (vev) of the $1/2$ BPS circular Wilson loop of ${\cal N}=4$ super Yang Mills in terms of color invariants, valid for any representation $R$ of any gauge group $G$. This expression allows us to discuss various exact relations among vevs in different representations. We also display the reduction of these color invariants to simpler ones, up to seventh order in perturbation theory, and verify that the resulting expression is considerably simpler for the logarithm of $\vev{W}_R$ than for $\vev{W}_R$ itself. We find that in the particular case of the symmetric and antisymmetric representations of SU($N$), the logarithm of $\vev{W}_R$ satisfies a quadratic Casimir factorization up to seventh order, and argue that this property holds to all orders. Finally, we derive the large $N$ expansion of $\vev{W}_R$ for an arbitrary, but fixed, representation of SU($N$), up to order 1/{$N^2$}.}

\begin{document}
\maketitle

\section{Introduction}
Wilson loops are among the fundamental operators in gauge theories. Nevertheless, when it comes to extracting physically interesting quantities, many of them are determined in terms of the logarithm of the vacuum expectation value (vev) of certain Wilson loops, and not the vevs themselves. For instance, the quark anti-quark static potential is determined from the logarithm of the vev of a rectangular Wilson loop. Similarly, the cusp anomalous dimension \cite{Polyakov:1980ca} is the logarithm of the properly regularized vev of a Wilson loop with a cusp, dependent on the boost parameter $\varphi$
\be
\vev{W_\varphi} \sim e^{\Gamma_{\text{cusp}}(\varphi) \ln \frac{\Lambda_{\text{UV}}}{\Lambda_{\text{IR}}}}
\label{cusp}
 \ee
The question then arises whether one can directly compute the logarithm of the vacuum expectation value of the Wilson loop, bypassing the computation of the vev of the Wilson loop itself. At the perturbative level, according to the non-Abelian exponentiation theorem \cite{Gatheral:1983cz,Frenkel:1984pz} (see \cite{White:2015wha} for a pedagogical review), for certain cases the answer is positive. One has to evaluate just a subset of the Feynman diagrams that would appear in the computation of the vev of the Wilson loop, with the proviso that each Feynman diagram carries now a modified color factor, and not the standard one assigned according to the ordinary Feynman rules. The application of the non-Abelian exponentiation theorem to the computation of the perturbative cusp anomalous dimension is discussed for QCD in \cite{Korchemsky:1991zp} and for ${\cal N}=4$ super Yang-Mills in \cite{Henn:2013wfa}.

In order to understand the content of the non-Abelian exponentiation theorem, it is very clarifying to consider Wilson loops $W_R$ in arbitrary representations $R$ of the gauge group $G$. The perturbative expansion of their vevs can then be written in terms of color invariants. These color invariants involve contractions of the fully symmetrized traces \cite{Cvitanovic:2008zz, vanRitbergen:1998pn}
\be
d_R^{a_1\dots a_n}=\frac{1}{n!} \hbox{ tr }\sum_{\sigma \in {\cal S}_n} T_R^{a_{\sigma(1)}}\dots T_R^{a_{\sigma (n)}}
\label{symtraces}
\ee	
where $T^a_R$ are the generators of the Lie algebra of the group $G$, in the representation $R$.\footnote{We follow the convention that the previous definition with no indices means $d_R= \text {tr}_R \, \mathbb{1}= \text{dim } R$. The Appendix contains our conventions for color invariants, a summary of techniques useful to evaluate them, and their evaluation for various representations and gauge groups.}    Some examples of color invariants are $d_R^{aabb}$ or $d_R^{abcd}d_A^{abcd}$. The non-Abelian exponentiation theorem implies that certain color invariants present in $\vev{W}_R$ are absent in $\ln \vev{W}_R$.

In this paper, we will consider the interplay of the non-Abelian exponentiation theorem and the evaluation of the vev of Wilson loops in ${\cal N}=4$ SYM, leaving the case of ${\cal N}=2$ SCFTs for future work. Before describing our results, we want to argue that this theorem provides evidence for a recently formulated conjecture \cite{Fiol:2015spa}. To present the conjecture, and our argument, it is necessary to introduce a couple of quantities that will also appear in the main body of the paper.

First, the Bremsstrahlung function $B$ associated to a heavy probe is defined \cite{Correa:2012at} from the small boost limit of the cusp anomalous dimension (\ref{cusp}),
\be
\Gamma_{\text{cusp}}(\varphi)=B\varphi^2 +{\cal O}(\varphi^4)
\label{bremss}
\ee
This coefficient determines a number of interesting properties of a heavy probe coupled to a conformal field theory: its energy loss by radiation \cite{Correa:2012at}, its momentum diffusion coefficient \cite{Fiol:2013iaa} and the change in entanglement entropy it causes  in a spherical region \cite{Lewkowycz:2013laa}. Since the cusp anomalous dimension satisfies the non-Abelian exponentiation theorem, so does the Bremsstrahlung function: only a subset of the most general color invariants will appear in its expansion. On the other hand, in any four-dimensional conformal field theory, the  two-point function of the stress-energy tensor and a straight Wilson line is determined by conformal invariance, up to a coefficient $h_W$ \cite{Kapustin:2005py}
\be
\frac{\vev{T^{00}(x)W}}{\vev{W}}=\frac{h_W}{|\vec x|^4}
\ee
This coefficient appears also in the two-point function of the stress-energy tensor and a circular Wilson loop. From this definition, there is no hint that $h_W$ should involve only a subset of color invariants. Nevertheless, for ${\cal N}=2$ SCFTs, these two coefficients are related as
\be
B=3 h_W
\label{bandhw}
\ee
This identity was first noticed to hold in ${\cal N}=4$ super Yang-Mills, by explicit computation \cite{Correa:2012at,Fiol:2012sg}; it was conjectured to hold for ${\cal N}=2$ SCFTs in \cite{Lewkowycz:2013laa, Fiol:2015spa} and recently proven in \cite{Bianchi:2018zpb}. However this identity is somewhat surprising in light of the previous comments. For arbitrary gauge group $G$ and representation $R$, $B$ can be expressed in terms of just a subset of color invariants. Why should that be the case also for $h_W$? In \cite{Fiol:2015spa}, it was further conjectured that for ${\cal N}=2$ SCFTs
\be 
h_W = \frac{1}{12\pi^2}\partial_b \ln \vev{W_b}  |_{b=1}
\label{hwfromwb}
\ee
where $\vev{W_b}$ is the vev of a circular Wilson loop in a squashed sphere of parameter $b$. This conjecture has been checked up to three loops  \cite{Fiol:2015spa, Gomez:2018usu}; we want to show that the non-Abelian exponentiation theorem provides evidence of this conjecture (\ref{hwfromwb}) by arguing that both sides of (\ref{hwfromwb}) involve at every order in perturbation theory the same subset of color invariants. On the one hand, given that (\ref{bandhw}) is now an established result \cite{Bianchi:2018zpb}, we know that $h_W$ involves that same subset of color invariants as $B$. On the other hand, by virtue of the non-Abelian exponentiation theorem, the perturbative expansion of $\ln \vev{W_b}$ involves also just the reduced set of color invariants implied by this theorem. What this argument doesn't prove is that the coefficients that appear in front of the color invariants in the expansions of both sides of  (\ref{bandhw}) also coincide; it doesn't address the non-perturbative validity of  (\ref{bandhw}) either. The same comments apply to similar relations between various Bremsstrahlung functions and logarithms of Wilson loops in 3d ABJM theories \cite{Bianchi:2014laa, Bianchi:2017svd, Bianchi:2017ozk, Bianchi:2018scb}.

After this detour, let's now describe the contents of the body of the paper. In this work we will focus on 1/2 BPS Wilson loops of ${\cal N}=4$ super Yang-Mills, and the quantities that can be obtained from these operators. Locally BPS Wilson loops of ${\cal N}=4$ super Yang-Mills depend on a representation $R$ of the gauge group $G$, and a spacetime contour ${\cal C}$
\be 
W_R[{\cal C}] =\frac{1}{\text{dim R}} \text{tr}_R {\cal P}\,
\text{exp} \left(i\int_{\cal C} (A_\mu \dot x^\mu +|\dot x|\Phi_i \theta^i)ds \right)
\ee
When the contour is a circle (in Euclidean signature) the vev of this Wilson loop can be computed by supersymmetric localization \cite{Pestun:2007rz} that reduces it to a Gaussian matrix model over the Lie algebra $\mathfrak{g}$
\be  
\vev{W}_R = \frac{1}{\text{dim }R} \frac{\int_{\mathfrak{g}} dM    \text{tr}_R e^M \,  e^{-\frac{1}{2g}\text{tr }M^2}}{\int_{\mathfrak{g}}dM
e^{-\frac{1}{2g}\text{tr }M^2}}
\label{mmlie}
\ee
The most common approach to tackle this type of matrix integrals is to first reduce the integral over the Lie algebra to an integral over a Cartan subalgebra $\mathfrak{h}$. This introduces a Jacobian, given by a Vandermonde determinant $\Delta(X)^2$,
\be  
\vev{W}_R = \frac{1}{\text{dim }R} \frac{\int_{\mathfrak{h}} dX \Delta(X)^2  \text{ tr}_R e^X \,  e^{-\frac{1}{2g}\text{tr }X^2}}{\int_{\mathfrak{h}}dX \Delta(X)^2\, e^{-\frac{1}{2g}\text{tr }X^2}}
\label{mmcartan}
\ee
Then one applies either the method of orthogonal polynomials at finite $N$, or the saddle point approximation at large $N$ (see \cite{Marino:2004eq} for a pedagogical review). This approach yields compact expressions for particular choices of $G$ and $R$, but obscures the generic structure. In the current work, we are not going to follow this approach. Instead, following recent works \cite{Billo:2017glv, Billo:2018oog} we will not restrict the integrals to a Cartan subalgebra $\mathfrak{h}$ as in (\ref{mmcartan}), but rather integrate over the full Lie algebra $\mathfrak{g}$, as in (\ref{mmlie}). At the technical level, the advantage is that the Vandermonde determinant is not generated, and the matrix integrals are truly trivial, since they are Gaussian. They can be carried out at once, for any $R$ and $G$, just applying Wick's theorem. At the conceptual level, the benefit of this approach is that the results obtained are in terms of color invariants. Our first result is that the vev of $W_R$ can be written in term of symmetrized traces (\ref{symtraces}), with pairwise contracted indices,
\be 
\vev{W}_R = \frac{1}{d_R}\sum_{k=0}^\infty d_R^{a_1 a_1\dots a_k a_k} \frac{1}{k!}\left(\frac{g_{\text{YM}}^2}{4}\right)^k
\label{introexact}
\ee
This expression gives the vev of 1/2 BPS circular Wilson loop for any representation $R$ of a gauge group $G$. It allows to discuss exact relations among vevs in different representations. For instance, if $R^t$ is the transpose representation of $R$ (in the sense of having Young diagrams transpose to each other), we will argue that
\be
\vev{W}_{R^t}(\lambda,N)=\vev{W}_R(\lambda,-N)
\ee
thus relating, for instance, vevs in the symmetric and the antisymmetric representations of SU($N$).
It is possible to take the logarithm of (\ref{introexact}), to obtain a closed expression for $\ln \vev{W}_R$, but this closed expression is of very little use; in particular, the non-Abelian exponentiation theorem is not manifest. On the other hand, the color invariants $d_R^{a_1 a_1\dots a_k a_k}$ in (\ref{introexact}) can be reduced to lower order color invariants. As it will be illustrated in the main body of the paper, this expansion is simpler for $\ln \vev{W}_R$ than for $\vev{W}_R$ itself: the only color invariants that appear in the perturbative expansion of $\ln \vev{W}_R$ at a given order are those that can't be written as products of color invariants that appear at lower orders of the perturbative expansion, thus providing an illustration of the non-Abelian exponentiation theorem.
	
The structure of the papers is as follows. In Section 2 we derive an exact expression for $\vev{W}_R$ in terms of color invariants, and present some exact relations among vevs of different representations. In Section 3 we study the large $N$ limit of $\vev{W}_R$ for arbitrary, but fixed, representations of SU($N$), up to order 1/\text{$N^2$}. In Section 4, we present the expansion of $\ln \vev{W}_R$ in terms of color invariants; we provide a diagrammatic interpretation of the expansion, and discuss some patterns present in the perturbative expansion. The Appendix contains our conventions for color invariants, a summary of the techniques we use to evaluate them, and tables of the evaluation of various color invariants.

\section{$\vev{W}_R$ in terms of color invariants}
In this section we revisit the evaluation of $\vev{W}_R$ for ${\cal N}=4$ super Yang Mills, for an arbitrary representation $R$ of a generic Lie algebra $G$. Thanks to supersymmetric localization \cite{Pestun:2007rz} this problem reduces to a Gaussian matrix model, and it has been solved exactly, for various choices of gauge group $G$ and representation $R$ \cite{Drukker:2000rr, Fiol:2013hna, Fiol:2014fla}.  As mentioned in the introduction, typically this is done by first reducing the matrix integral to an integral over the Cartan subalgebra, as in (\ref{mmcartan}). While this procedure allows to obtain compact expressions for $\vev{W}_R$ for some choices of $R$, this has to be done in a case by case basis, and it obscures the dependence on the choice of $G$ and $R$. Since in this work we are particularly interested in expressing $\vev{W}_R$ in terms of color invariants, we will follow a different route. We will instead carry out the integrals over the full Lie algebra. Specifically, 
\be  
\vev{W}_R = \frac{1}{d_R}\vev{\hbox{tr}_R \, e^{2\pi M}}=
\frac{1}{\text{dim }R} \frac{\int_{\mathfrak{g}} dM    \, \text{tr}_R e^{2\pi M} \,  e^{-\frac{8\pi^2}{g_{\text{YM}}^2}\text{tr }M^2}}{\int_{\mathfrak{g}}dM \,
e^{-\frac{8\pi^2}{g_{\text{YM}}^2}\text{tr }M^2}}
\label{winmm}
\ee
If we denote by $m^a$ the coefficients of the matrix $M$ in the Lie algebra, the two-point function in this Gaussian matrix model is
\be
\vev{m^a m^b}= \frac{g^2_\text{YM} }{8\pi^2} \delta^{ab} \hspace{1cm} a,b=1,\dots, d_A
\label{atwopoint}
\ee
To compute the vev of the normalized Wilson loop, we expand the exponent insertion in (\ref{winmm}), use the two-point function (\ref{atwopoint}) and apply Wick's theorem,

\be
\vev{W}_R =\frac{1}{d_R}\sum_{k=0}^\infty
\frac{(2\pi)^{2k}}{(2k)!} \vev{m^{a_1}\dots m^{a_{2k}}} \hbox{tr } T^{a_1}_R \dots T^{a_{2k}}_R=\frac{1}{d_R}\sum_{k=0}^\infty d_R^{a_1a_1\dots a_k a_k} \frac{g^k}{k!} 
\label{exactvev}
\ee
where $g=g^2_\text{YM}/4$ and $d_R^{a_1\dots a_k}$ are the symmetrized traces defined in (\ref{symtraces}). This expression for $\vev{W}_R^{{\cal N}=4}$ is exact - recall that there are no instanton corrections for $\vev{W}_R$ in ${\cal N}=4$ \cite{Pestun:2007rz} - and valid for any $G$ and any $R$. It encompasses and unifies all the known results for particular choices of $G$ and $R$ \cite{Drukker:2000rr,Fiol:2013hna, Fiol:2014fla}.

Now it is a matter of evaluating $d_R^{a_1 a_1\dots a_k a_k}$, the fully symmetrized traces (\ref{symtraces}) with pairwise contracted indices. At every order, the outcome is a combination of lower order color invariants, involving the original representation $R$, and the adjoint representation $A$. At low orders, it's easy enough to evaluate them by hand, using the techniques detailed in \cite{vanRitbergen:1998pn}. For instance,
\begin{align*}
d_R^{aa} & =  \text{tr } T_R^a T_R^a= c_R d_R \\
d_R^{aabb} & =  \frac{1}{3}\text{ tr}
\left(2T_R^aT_R^aT_R^bT_R^b+T_R^aT_R^bT_R^aT_R^b\right)=(c_R^2-\frac{1}{6}c_A c_R)d_R
\end{align*}
To push the evaluation to higher orders, we use FormTracer \cite{Cyrol:2016zqb}. Up to order $g_\text{YM}^{14}$ we obtain
\begin{multline}
\vev{W}_R=1+c_R g +\left(c_R^2-\frac{1}{6}c_Rc_A\right)\frac{g^2}{2!}+
\left(c_R^3-\frac{1}{2}c_R^2c_A+\frac{1}{12}c_Rc_A^2\right)\frac{g^3}{3!}+ \\
\left(c_R^ 4-c_R^3c_A+\frac{5}{12}c_R^2c_A^2-\frac{5}{72}c_Rc_A^3+\frac{1}{15}\frac{d_R^{abcd}d_A^{abcd}}{d_R}\right)\frac{g^4}{4!}+\\
\left(c_R^5-\frac{5}{3}c_R^4c_A+\frac{5}{4}c_R^3c_A^2-\frac{35}{72}c_R^2c_A^3+\frac{35}{432}c_Rc_A^4+(\frac{1}{3}c_R-\frac{2}{9}c_A)\frac{d_R^{abcd}d_A^{abcd}}{d_R}
+\frac{1}{90}c_R\frac{d_A^{abcd}d_A^{abcd}}{d_A} \right)\frac{g^5}{5!} \\
+\left(-\frac{35}{288} c_R c_A^5 + \frac{35}{48} c_A^4 c_R^2-\frac{35}{18} c_A^3 c_R^3+\frac{35}{12} c_A^2 c_R^4 - \frac{5}{2} c_A c_R^5 +  c_R^6    +\frac{1}{10} \frac{d_R^{abcd}d_A^{cdef}d_A^{efab}}{d_R} \right.  \\ 
\left. + ( \frac{1}{15}c_R^2-\frac{11}{180} c_R c_A ) \frac{d_A^{abcd} d_A^{abcd}}{d_A}+  (c_R^2 - \frac{3}{2}c_A c_R +\frac{11}{18} c_A^2) \frac{d_R^{abcd}d_A^{abcd}}{d_R} -\frac{8}{63} \frac{d_R^{abcdef} d_A^{abcdef}}{d_R} \right) \frac{g^6}{6!}+\\
+ \left(c_R^7 - \frac{7}{2} c_A c_R^6 + \frac{35}{6} (c_A^2 c_R^5 - c_A^3 c_R^4) + \frac{175}{48} c_A^4 c_R^3 - \frac{385}{288} c_A^5 c_R^2 + \frac{72757}{326592} c_R c_A^6 \right. +  \\
(\frac{7}{3}c_R^3-\frac{35}{6}c_R^2c_A+\frac{21}{4}c_Rc_A^2 -\frac{91}{54}c_A^3) \frac{d_R^{abcd} d_A^{abcd}}{d_R}+(\frac{7}{30}c_R^3-\frac{7}{15}c_R^2 c_A+\frac{817}{3240}c_R c_A^2)\frac{d_A^{abcd} d_A^{abcd}}{d_A}+ \\
\frac{691}{18900} c_R \frac{d_A^{abcd} d_A^{cdef} d_A^{efab}}{d_A} -\frac{80}{27} \frac{d_A^{abcdef} d_R^{abcg} d_A^{defg}}{d_R} + \frac{14}{45} \frac{d_R^{abcdef} d_A^{abcg} d_A^{defg}}{d_R}+\\ 
\left. \frac{7}{10}(c_R-c_A) \frac{d_R^{abcd} d_A^{cdef} d_A^{efab}}{d_R} + \frac{8}{9}(c_A-c_R) \frac{d_R^{abcdef} d_A^{abcdef}}{d_R} \right) \frac{g^7}{7!}+\dots  
\label{vevexpan}
\end{multline}
By construction, every color invariant in this expansion involves an even number of indices. Since $d_A^{a_1\dots a_k}=0$ for $k$ odd, for every color invariant the adjoint representation contributes an even number of indices, and thus the representation $R$ also contributes an even number. Up to the order computed, no color invariants involving $d_R^{a_1\dots a_k}$ with an odd number of indices appear in the expansion. They would necessarily involve more than one $d_R^{a_1\dots a_k}$, e.g. $d_R^{abc}d_R^{ade}d_A^{bcde}$. It is not clear to us whether such color invariants will appear at higher orders.

The reader that feels intimidated by the expansion (\ref{vevexpan}) might find some comfort in the fact that, as we will show in Section 4, the perturbative expansion of $\ln \vev{W}_R$ is considerably simpler. 

Besides the possibility of evaluating $\vev{W}_R$ order by order in $g_\text{YM}$ for all $G$ and $R$ at once, the result (\ref{exactvev}) allows to derive some general exact relations among vevs of 1/2 BPS Wilson loops in different representations. The first identity that we will point out is rather obvious. For a generic representation $R$, recall that the complex conjugate representation $\bar R$ of $R$ has generators $T_{\bar R}=-T^t_{R}$. Then, since dim $R$ = dim $\bar R$, it follows that
\be
\vev{W}_{\bar R}=\vev{W}_{R}
\label{conjugaterel}
\ee
As an illustration of this equality, we have that $\vev{W}_{A_k}^{\text{SU}(N)}=\vev{W}_{A_{N-k}}^{\text{SU}(N)}$, an identity that is readily seen to hold in the explicit results of  \cite{Fiol:2013hna}. 

A less trivial relation involves representations of classical Lie groups with transposed Young diagrams. For instance, irreducible representations of SU($N$) are labelled by Young diagrams, and exchanging symmetrization and antisymmetrization of indices amounts to transposing the Young diagram. It is known \cite{Cvitanovic:1982bq, Cvitanovic:2008zz} that under this operation, color invariants change as $N\rightarrow -N$, up to an overall sign. For a representation $R$ whose Young diagram has $k$ boxes, the overall sign is
\be 
d_{R^t}^{a_1a_1\dots a_m a_m}(N)=(-1)^{k+m}d_R^{a_1a_1\dots a_m a_m}(-N)
\ee
In particular, dim $R^t(N)=(-1)^k$ dim $R(-N)$, as one can check for SU($N$) using (\ref{dimofrep}). Since we are considering normalized Wilson loops, divided by dim $R$, the $(-1)^k$ cancels in (\ref{exactvev}). The remaining $(-1)^m$ can be absorbed by expanding the vev in powers of the 't Hooft coupling $\lambda=g^2_{\text{YM}}N$ instead of in powers of $g^2_{\text{YM}}$. Overall, we arrive at the relation
\be
\vev{W}_{R^t}(\lambda,N)=\vev{W}_R(\lambda,-N)
\label{transposerel}
\ee
In the next section, we will provide an alternative derivation of this identity for SU($N$). As a first illustration, a particular example of this identity is the relation 
\be
\vev{W}_F^{\text{Sp}(N)}(N)=\vev{W}_F^{\text{SO}(2N)}(-N)
\ee
found in \cite{Fiol:2014fla}. Moreover, (\ref{transposerel}) implies that the vevs of Wilson loops in the symmetric and antisymmetric representations of SU($N$) satisfy
\be
\vev{W}_{S_k}(\lambda,N)=
\vev{W}_{A_k}(\lambda,-N)
\label{relsymanti}
\ee
since the Young diagrams of the $k-$symmetric and $k-$antisymmetric representations are transpose of each other. Okuyama \cite{Okuyama:2018aij} recently found evidence for this particular consequence of the identity in eq. (\ref{transposerel}).\footnote{To compare our identity (\ref{relsymanti}) with the one in \cite{Okuyama:2018aij}, note that the Wilson loops in \cite{Okuyama:2018aij} are not normalized by the dimension of the representation. If the Young diagram associated with the representation $R$ has $k$ boxes, dim $R^t(N)=(-1)^k$ dim $R(-N)$, so normalizing the Wilson loop by dim $R$ introduces an additional $(-1)^k$ factor in the relation, proving the equivalence of the result in \cite{Okuyama:2018aij} and ours.} To illustrate the relation in eq. (\ref{relsymanti}) we evaluate (\ref{exactvev}) for $G=\text{SU}(N)$ and $R=S_k,A_k$ up to seventh order in $\lambda$, applying methods explained in the Appendix. For compactness, we actually display the perturbative expansion of $\ln \vev{W}_{S_k/A_k}$, with the upper signs corresponding to the symmetric representation, and the lower signs to the antisymmetric one,
\begin{multline}
\ln \vev{W}_{S_k/A_k}= \frac{k(N\pm k)(N\mp 1)}{2N^2} \left(\frac{\lambda}{4}-\frac{1}{12}\left(\frac{\lambda}{4}\right)^2+\frac{1}{72}\left(\frac{\lambda}{4}\right)^3 +\frac{-4N^2\mp N+k(k\pm N)}{1440 N^2} \left(\frac{\lambda}{4}\right)^4 + \right. \\
\frac{13N^2\pm 10N-10k(k\pm N)+3}{21600N^2} \left(\frac{\lambda}{4}\right)^5 -
\left(\frac{11}{112}+\frac{43(\pm N-k(k\pm N))}{280N^2}+\frac{109}{1008 N^2}+\right. \\
 \left . \frac{\pm 73N-113Nk(N\pm k)+20k^2(N\pm k)^2}{ 2520N^4}\right)\frac{1}{6!}\left(\frac{\lambda}{4}\right)^6 
+\left(\frac{647}{20321280}\pm \frac{89}{1036800 N} \right. \\
+\frac{10501}{101606400 N^2}\pm\frac{17}{268800N^3}+\frac{197}{12700800N^4} \mp \frac{5471 k}{65318400 N}
-\frac{19}{268800 N^2}-\frac{4499k^2}{65318400 N^2} \\
\left. \left. \mp \frac{k}{44800 N^3}\mp \frac{19 k^2}{268800 N^3} \pm\frac{k^3}{33600 N^3} -\frac{k^2}{44800 N^4}+\frac{k^4}{67200 N^4} \right)\left(\frac{\lambda}{4}\right)^7 \right)+\dots
\label{pertskak}
\end{multline}
Notice that, at least up to order $g_\text{YM}^{14}$, all the coefficients factorize, and have a common factor that happens to be essentially the quadratic Casimir $c_{S_k/A_k}$,
\be
c_{S_k/A_k}=\frac{k(N\pm k)(N\mp 1)}{2N}
\ee
This factorization is unexpected and, as the next example shows, it does not happen for generic representations. In the next section we will discuss this factorization in more detail, and argue that for
$\ln \vev{W}_{S_k/A_k}$ it holds to all orders.

Another implication of  the identity (\ref{transposerel}) is that if $R$ is a SU($N$) representation with a self-transpose Young diagram, $\vev{W}_R(\lambda,N)$ admits a $1/N^2$ rather than the more general $1/N$ expansion. A first illustration of this point is the fact that $\vev{W}^{\text{SU}(N)}_{\ydiagram{1}}$ has a $1/N^2$ expansion. As a second illustration of this point, we display the perturbative expansion of $\ln \vev{W}^{\text{SU}(N)}_{\ydiagram{2,1}}$ up to seventh order in $\lambda$, showing that every coefficient has a $1/N^2$ expansion,
\begin{multline}
\ln \vev{W}^{\text{SU}(N)}_{\ydiagram{2,1}}= 
\frac{3(N^2-3)}{2N^2}\frac{\lambda}{4}-\frac{N^2-3}{8N^2}\left(\frac{\lambda}{4}\right)^2+
\frac{N^2-3}{48N^2}\left(\frac{\lambda}{4}\right)^3 
+\frac{-4N^4+19N^2-27}{960 N^4}\left(\frac{\lambda}{4}\right)^4 \\
+\frac{13N^4-106N^2+261}{14400N^4}\left(\frac{\lambda}{4}\right)^5
+\frac{-495 N^6+6796N^4-23269N^2+9720}{2419200 N^6}\left(\frac{\lambda}{4}\right)^6 \\
+\frac{3235 N^6-71360 N^4+310273 N^2 -268572}{67737600 N^6} \left(\frac{\lambda}{4}\right)^7 +\dots
\label{perthook}
\end{multline}

%It is worth pointing out that this last implication does not work the %other way around. For instance, for $G=SU(N)$ and $R$ the adjoint %representation, the Young diagram is not self-transpose, and yet %$\vev{W}_{Adj}$ has a $1/N^2$ expansion.

While we are discussing identities (\ref{conjugaterel}) and (\ref{transposerel}) for 1/2 BPS circular Wilson loops of ${\cal N}=4$ SYM, since they are mostly based on group theoretic properties of the color invariants, we expect that similar identities hold in more generic theories, for other observables defined in terms of a representation $R$ of a classical Lie group $G$.

Equations (\ref{conjugaterel}) and (\ref{transposerel}) are exact relations, valid for finite $\lambda$ and $N$. When the gauge group has a classical Lie algebra (and therefore a large $N$ gravity dual), these exact relations have implications for the holographic dual. In particular, let's comment briefly on the implications of $\vev{W}^{\text{SU}(N)}_R$ having a $1/N^2$ expansion when $R$ is a representation with a self-transpose Young diagram. 

In the probe limit, the holographic dual to a Wilson loop operator with an arbitrary Young diagram is a system of D3 and D5-branes in IIB \cite{Gomis:2006sb,Drukker:2005kx, Yamaguchi:2006tq}. Considering the transpose representation amounts to exchanging D3 and D5 branes. The identity (\ref{transposerel}) implies that in the particular case when the D-brane system is invariant under the exchange of D3 and D5 branes, corrections have a $1/N^2$ expansion. If we keep increasing the size of the Young diagram, the correct dual gravitational description eventually is in terms of bubbling geometries, half-BPS solutions of IIB supergravity, fully described in \cite{DHoker:2007mci}. The representation $R$ is geometrically encoded in a hyperelliptic curve, and a self-transpose Young diagram corresponds to hyperelliptic curves with an additional $\bZ_2$ symmetry. Again, our results imply that corrections to the supergravity action evaluated on these backgrounds have $1/N^2$ as expansion parameter, instead of $1/N$. It would be interesting to check these predictions on the various regimes of the holographic dual.

\section{Large $N$ expansion of $\vev{W}_R^{\text{SU}(N)}$}
In this section we expand the vev of the unnormalized 1/2 BPS Wilson loop for a generic but fixed representation of SU($N$) in the large $N$ limit. We will obtain the leading term, the $1/N$ and the $1/N^2$ corrections. We do so for a fixed representation, i.e. we do not consider the interesting case where the number of boxes in the Young diagram of the representation scales with N. For recent work in that direction see \cite{Chen-Lin:2016kkk,Gordon:2017dvy, Okuyama:2017feo,CanazasGaray:2018cpk}.

Let $R$ be an arbitrary irreducible representation of SU($N$), whose associated Young diagram has $k$ boxes. This Young diagram is also associated to an irreducible representation $R$ of the symmetric group ${\cal S}_k$. In the large $N$ limit, the vev of the unnormalized (i.e. not divided by the dimension of R) Wilson loop grows like $\vev{W}_R \sim N^k$ \footnote{Since dim $R$ $\sim N^k$ also, the vev of the normalized Wilson loop has a $N^0$ leading term.}. 

We are now going to write $\vev{W}_R$ as a sum of $n-$point functions of multiply-wound Wilson loops, with $n\leq k$. To do so, we need to recall some basic facts about the symmetric group ${\cal S}_k$ \cite{sagansymmetric}. A permutation $\pi \in {\cal S}_k$ is of cycle type $(1^{m_1},2^{m_2},\dots, k^{m_k})$ if it has $m_j$ cycles of length $j$. Two permutations of ${\cal S}_k$ are in the same conjugacy class if and only if have the same cycle type. Conjugacy classes of ${\cal S}_k$ are labelled by partitions of $k$, or equivalenty by Young diagrams with $k$ boxes: the conjugacy class $\lambda=(1^{m_1},2^{m_2},\dots, k^{m_k})$ corresponds to a diagram with $m_j$ rows of $j$ boxes. Finally, if we define $z_\lambda= 1^{m_1} m_1! 2^{m_2} m_2! \dots k^{m_k} k!$, the number of elements in the conjugacy class $\lambda$ is $k!/z_\lambda$.

\begin{figure}
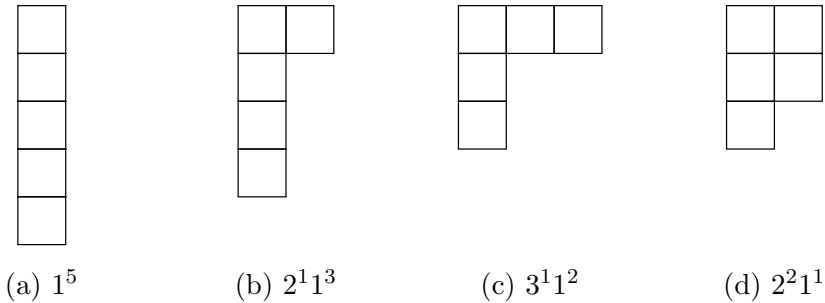

\centering
\begin{subfigure}{0.2\textwidth}
  \centering
  \ytableausetup{boxsize=normal}
  \ytableausetup{nosmalltableaux}
    \ydiagram{1,1,1,1,1} 
    \caption{$1^5$}    
  \end{subfigure}
  \begin{subfigure}{0.2\textwidth}
    \centering
    \ytableausetup{boxsize=normal}
    \ytableausetup{nosmalltableaux}
    \ydiagram{2,1,1,1,0}
    \caption{$2^1 1^3$}
  \end{subfigure}
  \begin{subfigure}{0.2\linewidth}
    \centering
    \ydiagram{3,1,1,0,0}
    \caption{$3^1 1^2$}
  \end{subfigure}
  \begin{subfigure}{0.2\linewidth}
    \centering
      \ydiagram{2,2,1,0,0}
      \caption{$2^2 1^1$}
  \end{subfigure}
\caption{The large $N$ expansion of $\vev{W}_R^{\text{SU}(N)}$ for any fixed $R$ can be computed up to order $1/N^2$, in terms of four correlators of multiply-wound Wilson loops. Each correlator of multiply-wound Wilson loops has its own Young diagram, and the four relevant ones are displayed in this figure. They are shown for the particular example of an arbitrary irreducible representation $R$ with $k=5$ boxes in its Young diagram. The first one contributes at leading order, and at $1/N^2$ order. The second one contributes at $1/N$ order. The last two ones contribute at $1/N^2$ order.}
\label{fouryoung}
\end{figure}

\ytableausetup{boxsize=5pt}

We are now ready to write $\vev{W}_R$ as a sum of $n-$point functions of multiply-wound Wilson loops in the fundamental representation. Denoting by $W(n)$ the n-times wound Wilson loop, by virtue of Frobenius theorem \cite{sagansymmetric},
\be 
\vev{W}_R = \sum_{\lambda} \frac{\chi^r(\lambda)}{z_\lambda} 
\vev{W(1)^{m_1}\dots W(k)^{m_k}}
\label{frobenius}
\ee
where $\chi^r(\lambda)$ is the character of $R$ evaluated in the conjugacy class $\lambda$. This relation is exact, valid for finite $N$. The sum in (\ref{frobenius}) is over conjugacy classes of ${\cal S}_k$, so there are $p(k)$ terms, the number of partitions of $k$. However in the large $N$ limit, only a few of these terms contribute to the leading behavior and the first subleading corrections. In fact, we will argue that to compute the first three terms in the large $N$ expansion of $\vev{W}_R$, one needs to consider only four terms in the sum (\ref{frobenius}).

Let's now recall a couple of properties of the $n-$point functions $\vev{W(1)^{m_1}\dots W(k)^{m_k}}$.  Large $N$ factorization implies that the leading behavior is given by $N^{\sum_j m_j}$; notice that $\sum_j m_j$ is the number of rows of the corresponding Young diagram. Furthermore, all these correlators have a $1/N^2$ expansion. These two properties allow us to give a different derivation of (\ref{transposerel}) for SU($N$), or more precisely, its formulation for unnormalized Wilson loops,
\be  
\vev{W}_{R^t}(\lambda,N)=(-1)^k \vev{W}_R(\lambda,-N)
\label{unnormtrans}
\ee
The argument goes as follows. If $R$ is an irreducible representation of ${\cal S}_k$, $r^t=r\otimes \text{sgn}$ is also an irreducible representation, and their Young diagrams are transpose of each other. We then have $\chi^{r^t}(\lambda)=\text{sgn }\lambda \, \chi^r(\lambda)$. The sign of a permutation can be easily read off from its Young diagram,
\be 
\text {sgn } \lambda=(-1)^{k-\sum_j m_j}
\label{signofperm}
\ee
where $k$ is the total number of boxes and $\sum_j m_j$ is the number of rows. In other words, the exponent is the total number of boxes not in the first column. On the other hand, according to the two properties explained above, $\vev{W(1)^{m_1}\dots W(k)^{m_k}}$ picks a sign $(-1)^{\sum_j m_j}$ under $N\rightarrow -N$. Plugging these two results into (\ref{frobenius}) yields the relation (\ref{unnormtrans}). 

Let's discuss now the correlators of multiply wound Wilson loops that contribute to the leading terms of the large $N$ expansion of $\vev{W}_R$. There is just one $\lambda$ whose Young diagram has $k$ rows, the vertical column, see figure (\ref{fouryoung}). This is the only $n-$point function contributing to the leading term, of order $N^k$, and because it has a $1/N^2$ expansion, it also contributes at order $N^{k-2}$, but not at order $N^{k-1}$. For its conjugacy class, $z_{1^k}=k!$. There is also just one $(k-1)$-point function contributing at order $N^{k-1}$, the one corresponding to the Young diagram where the $k$ boxes are distributed in $k-1$ rows, $(2^1 1^{k-2})$, see figure (\ref{fouryoung}). For its conjugacy class $z_{2^1 1^{k-2}}= 2 (k-2)!$. At order $1/N^2$, there are subleading contributions from $\vev{W(1)^k}$, and also leading contributions from two $(k-2)$-point functions, $\vev{W(3)W(1)^{k-3}}$ and $\vev{W(2)^2W(1)^{k-4}}$, see figure (\ref{fouryoung}) . All in all,
\begin{multline}
\vev{W}_R=\frac{\chi^r(1^k)}{k!}\vev{W(1)^k}+\frac{\chi^r(2^1 1^{k-2})}{2 (k-2)!}\vev{W(2)W(1)^{k-2}}+\frac{\chi^r(3^1 1^{k-3})}{3 (k-3)!}\vev{W(3)W(1)^{k-3}}+\\
\frac{\chi^r(2^2 1^{k-4})}{8 (k-4)!}\vev{W(2)^2 W(1)^{k-4}}+{\cal O}(N^{k-3})
\end{multline}
To compute the leading contributions to the vevs, we use that in the large $N$ limit, the $n-$point functions of the Gaussian matrix model factorize, and in the planar limit \cite{Erickson:2000af},
\be 
\vev{\frac{1}{N}W(n)} \rightarrow \frac{2}{n\sqrt{\lambda}}I_1(n\sqrt{\lambda})
\ee
where $I_1(\sqrt{\lambda})$ is the modified Bessel function. Let's consider finally the subleading contributions of $\vev{W(1)^k}$, that contribute at order $1/N^2$. To do so, it is convenient to write $\vev{W(1)^k}$ in terms of connected correlators. At this order the relevant terms are
\be 
\vev{W(1)^k}=\vev{W(1)}^k+{k \choose 2} \vev{W(1)}^{k-2} \vev{W(1)W(1)}_c +\dots
\ee
The dots correspond to more connected diagrams, which don't contribute at $1/N^2$ order. We see that $1/N^2$ contributions can come from two types of diagrams: first, from diagrams with $k$ disconnected pieces, $k-1$ planar ones and a non-planar one; second, from planar diagrams with $k-1$ disconnected pieces ($k-2$ of them are $1-$point functions, the last one is a connected 2-point function). The first contribution is obtained expanding the exact result of \cite{Drukker:2000rr}
\begin{equation*}
\vev{\frac{1}{N}W(1)}^k=\left(\frac{2 I_1(\sqrt{\lambda})}{\sqrt{\lambda}}+\frac{\lambda I_2(\sqrt{\lambda})}{48 N^2}+\dots\right)^k=
\frac{2^k I_1(\sqrt{\lambda})^k}{\lambda^{\frac{k}{2}}}+
k\frac{2^k I_1(\sqrt{\lambda})^{k-1} I_2(\sqrt{\lambda})}{96 \lambda^{\frac{k-3}{2}}} \frac{1}{N^2}+\dots
\end{equation*}
For the second contribution we need the leading term of the connected two-point function of Wilson loops $\vev{W(1)W(1)}_c $ \cite{Akemann:2001st},
\be 
\vev{W(1) W(1)}_c= \frac{\sqrt{\lambda} I_0(\sqrt{\lambda}) I_1(\sqrt{\lambda})}{2}+\dots
\ee
All in all, the vev of the unnormalized Wilson loop has the following 1/$N$ expansion,
\begin{multline}
\vev{W}_R^{\text{U}(N)}=\frac{\chi^r(1^k)}{k!} \left(\frac{2}{\sqrt{\lambda}}I_1(\sqrt{\lambda})\right)^k N^k+
\frac{\chi^r(2^1 1^{k-2})}{4 (k-2)!} \left(\frac{2}{\sqrt{\lambda}}\right)^{k-1} I_1(\sqrt{\lambda})^{k-2}I_1(\sqrt{4\lambda}) N^{k-1}+\\
\left(\frac{\chi^r(1^k)}{(k-1)!} \frac{2^k I_1(\sqrt{\lambda})^{k-1}}{16 \lambda^{(k-3)/2}}
\left( \frac{I_2(\sqrt{\lambda})}{6}+(k-1)I_0(\sqrt{\lambda})\right)+  
\frac{\chi^r(3^1 1^{k-3})}{3 (k-3)!}\frac{2^k I_1(\sqrt{\lambda})^{k-3} I_1(\sqrt{9 \lambda})}{ 12 \lambda^{(k-2)/2}}+ \right. \\
\left.\frac{\chi^r(2^2 1^{k-4})}{8(k-4)!} \frac{2^k I_1(\sqrt{\lambda})^{k-4} I_1(\sqrt{4 \lambda})^2}{16 \lambda^{(k-2)/2}}\right)N^{k-2}+\dots
\label{largenresult}
\end{multline}

We are now going to check that the general expansion (\ref{largenresult}) reproduces the explicit computations presented in the previous section. In order to make a detailed comparison, there are a couple of factors to take into account. The first one is that in the rest of the paper, the vevs are for SU($N$) and not for U($N$). This is not relevant for the leading term, but it affects the subleading terms. For the vev of $1/2$ BPS Wilson loop in a representation $R$ whose Young diagram has $k$ boxes, they are related by
\be
\vev{W}_R^{\text{SU}(N)}=e^{-\frac{\lambda k^2}{8N^2}}\vev{W}_R^{\text{U}(N)}
\ee
Since $k$ is fixed (it does not scale with $N$), this introduces a correction at order $1/N^2$. The other issue is that in this section, unlike in the rest of the paper, we have been considering Wilson loops not normalized by the dimension. However, since we will compare the generic result with explicit computations of $\ln \vev{W}_R$, the dimension only contributes as a coupling-independent additive constant.

As a first check, let's consider the case of the $S_k$ and $A_k$ representations of SU($N$). The corresponding representations of the symmetric group ${\cal S}_k$ are the trivial and the sign representations: $k^1$ and $1^k$, respectively. Since these are one-dimensional representations of ${\cal S}_k$, their characters coincide with the representation elements: $\chi^{k}(\pi)=1$, and $\chi^{1^k} (\pi)= \text{sgn } \pi$. The signs of the four relevant permutations can be computed using (\ref{signofperm}) and consulting the figure (\ref{fouryoung}). Applying then the formula (\ref{largenresult}) to $S_k/A_k$ we obtain, up to a coupling-independent constant,
\begin{multline}
\ln \vev{W}^{\text{SU}(N)}_{S_k/A_k}=  k \ln \frac{I_1(\sqrt{\lambda})}{\sqrt{\lambda}} \pm \frac{k(k-1)}{8}\frac{\sqrt{\lambda}I_1(\sqrt{4\lambda})}{I_1(\sqrt{\lambda})^2}\frac{1}{N}+ 
\left(-\frac{k^2}{8}\lambda+\frac{k\lambda^{3/2}I_2(\sqrt{\lambda}) }{96I_1(\sqrt{\lambda})}+\right.\\
\left. +\frac{k(k-1)\lambda^{3/2}I_0(\sqrt{\lambda}) }{16I_1(\sqrt{\lambda)}}
+\frac{k(k-1)(k-2) \lambda I_1(\sqrt{9\lambda})}{36 I_1(\sqrt{\lambda})^3}
-\frac{k(k-1)(2k-3) \lambda I_1(\sqrt{4\lambda})^2}{64 I_1(\sqrt{\lambda})^4}\right)\frac{1}{N^2}
\end{multline}
The $1/N$ correction vanishes for $k=1$, as it had to, since then the Wilson loop admits a $1/N^2$ expansion. This expression correctly reproduces the leading, $1/N$ and $1/N^2$ terms of the first orders computed in (\ref{pertskak}).

As a second check of the general result (\ref{largenresult}), consider the representation $\ydiagram{2,1}$. Its character evaluated on the relevant conjugacy classes is $\chi^{\ydiagram{2,1}}(\ydiagram{1,1,1})=2$, $\chi^{\ydiagram{2,1}}(\ydiagram{2,1})=0$, $\chi^{\ydiagram{2,1}}(\ydiagram{3})=-1$. The evaluation of (\ref{largenresult}) is then 
\be
\ln \vev{W}_{\ydiagram{2,1}}^{\text{SU}(N)}= 3 \ln \frac{I_1(\sqrt{\lambda})}{\sqrt{\lambda}}+
\left(-\frac{9}{8}\lambda+\frac{3\lambda^{3/2}I_0(\sqrt{\lambda})}{8 I_1(\sqrt{\lambda})}
+\frac{\lambda^{3/2}I_2(\sqrt{\lambda})}{32 I_1(\sqrt{\lambda})}
-\frac{\lambda I_1(\sqrt{9 \lambda})}{12 I_1(\sqrt{\lambda})^3}\right)\frac{1}{N^2}+\dots
\ee
which correctly reproduces the explicit computations displayed in (\ref{perthook}).

\section{Logarithm of $\vev{W}_R$}
In Section 2 we have obtained a formula (\ref{exactvev}) for the vev of the $1/2$ BPS Wilson loop, for arbitrary $G$ and $R$. For many applications, we are actually interested in  $\ln \vev{W}_R$. In this section we will obtain a closed expression for $\ln \vev{W}_R$ and discuss its perturbative expansion.

From (\ref{exactvev}) it is possible to write the power series for the logarithm of $\vev{W}_R$, in terms of partial Bell polynomials $B_{n,k}$. Defining $f_k=d_R^{a_1a_1\dots a_k a_k}/N_R$
\be
\ln \vev{W}_R =\sum_{k=1}^\infty \frac{g^k}{k!}\sum_{j=1}^k (-1)^{j-1} (j-1)! B_{k,j}(f_1,f_2,\dots,f_{k-j+1})
\label{logwbell}
\ee
As an application, using the result of \cite{Correa:2012at}, we obtain a closed formula for the Bremsstrahlung function (\ref{bremss}) of any 1/2 BPS particle, for generic $G$ and $R$
\be
B_R(\lambda,N)=\frac{1}{2\pi^2}\lambda \frac{\partial \ln \vev{W}_R}{\partial \lambda}=\frac{1}{2\pi^2} \sum_{k=1}^\infty \frac{g^k}{(k-1)!}\sum_{j=1}^k (-1)^{j-1} (j-1)! B_{k,j}(f_1,f_2,\dots,f_{k-j+1})
\label{closedb}
\ee
Taking into account the $B=3h_W$ relation (\ref{bandhw}), this also gives an expression for the coefficient $h_w$ appearing in the two-point function of the $1/2$ BPS Wilson loop and the stress-energy tensor, for arbitrary gauge group $G$ and representation $R$. 

While (\ref{logwbell}) is a closed expression for $\ln \vev{W}_R$, valid for any $G$ and any $R$, it is extremely inefficient, and it obscures the fact that the perturbative expansion of $\ln \vev{W}_R$ is actually simpler than that of $\vev{W}_R$. To make this point manifest, let's compute $\ln \vev{W}_R$ from eq. (\ref{vevexpan}), up to order $g_{\text{YM}}^{14}$, 
\begin{multline}
\ln \vev{W}_R=c_Rg- \frac{1}{6}c_Rc_A \frac{g^2}{2!}+\frac{1}{12}c_Rc_A^2 \frac{g^3}{3!}+
\left(-\frac{5}{72}c_Rc_A^3+\frac{1}{15}\frac{d_R^{abcd}d_A^{abcd}}{N_R}\right)\frac{g^4}{4!}+\\
\left( \frac{35}{432}c_Rc_A^4-\frac{2}{9}c_A\frac{d_R^{abcd}d_A^{abcd}}{N_R}
+\frac{1}{90}c_R\frac{d_A^{abcd}d_A^{abcd}}{d_A} \right)\frac{g^5}{5!}+ \\
 {\left(-\frac{35}{288}c_R c_A^5 + \frac{11}{18} c_A^2 \frac{d_R^{abcd}d_A^{abcd}}{d_R} - \frac{11}{180} c_A c_R \frac{d_A^{abcd}d_A^{abcd}}{d_A}  +\frac{1}{10} \frac{d_R^{abcd}d_A^{cdef}d_A^{efab}}{d_R} - \frac{8}{63} \frac{d_R^{abcdef}d_A^{abcdef}}{d_R} \right) \frac{g^6}{6!}}+ \\ 
\left( \frac{72757}{326592} c_R c_A^6  -\frac{91}{54} c_A^3 \frac{d_R^{abcd} d_A^{abcd}}{d_R} + \frac{817}{3240}c_A^2 c_R \frac{d_A^{abcd} d_A^{abcd}}{d_A} -\frac{7}{10} c_A \frac{d_R^{abcd} d_A^{cdef} d_A^{efab}}{d_R} \right. \\
\left.
+ \frac{691}{18900} c_R  \frac{d_A^{abcd} d_A^{abef} d_A^{cdef}}{d_A} -\frac{8}{27} \frac{d_A^{abcdef} d_R^{abcg} d_A^{defg}}{d_R}+\frac{14}{45} \frac{d_R^{abcdef} d_A^{abcg} d_A^{defg}}{d_R} +\frac{8}{9} c_A \frac{d_R^{abcdef} d_A^{abcdef}}{d_R} \right) \frac{g^7}{7!}+\dots
\label{logofwilson}
\end{multline}
Comparing with (\ref{vevexpan}), we see that many color invariants present in the expansion of $\vev{W}_R$ are absent in the expansion of $\ln \vev{W}_R$. For instance, there are no color invariants in (\ref{logofwilson}) involving $c_R^k$ with $k\geq 2$. This simpler structure is a consequence of the non-Abelian exponentiation theorem \cite{Gatheral:1983cz, Frenkel:1984pz}: at every order in perturbation theory, the only color invariants that can appear in $\ln \vev{W}_R$ are the ones that can't be written as products of color invariants that appear at lower orders in the perturbative expansion of $\vev{W}_R$. So in practice, to obtain the expansion of $\ln \vev{W}_R$ in terms of color invariants, it is more efficient to expand $\vev{W}_R$ as in (\ref{vevexpan}) and then discard by hand the terms that involve products of lower order color invariants.

\subsection{Casimir factorization}
In Section 2, we noticed that the evaluation of the perturbative expansion (\ref{logofwilson}) of $\ln \vev{W}_R$ for the case of the symmetric and antisymmetric representations of SU($N$), eq. (\ref{pertskak}), showed an unexpected pattern up to the computed order, that we will refer to as Casimir factorization (not to be confused with the Casimir scaling hypothesis, as we discuss below).  

On general grounds, the coefficients at every order in $\lambda$ in $\ln \vev{W}_R$ are polynomials in $1/N$. Equation (\ref{pertskak}) shows that up to at least order $\lambda^7$, these coefficients factorize (as polynomials in $1/N$) with a universal factor, the quadratic Casimir divided by $N$, which is also quadratic polynomial in $1/N$. We refer to this feature as Casimir factorization. We will now argue that Casimir factorization of $\ln \vev{W}_{S_k/A_k}$ holds to all orders
\be
\ln \vev{W}_{S_k/A_k} \overset{?}{=} \frac{c_{S_k/A_k}}{N} \, f_{S_k/A_k}(\lambda,N)
\label{modcasimir}
\ee
where $f_{S_k/A_k}$ is such that at every order in $\lambda$ the coefficient is a $k-$dependent {\it polynomial} in $1/N$. Recall that
\be 
c_{S_k/A_k}=\frac{k(N\pm k)(N\mp 1)}{2N}
\ee
so if we argue that at every order the coefficients of $\ln{W}_{S_k/A_k}$ are divisible by $(N\pm k)$ and $(N\mp 1)$, we are done. First, because of the relation $\vev{W}_{A_k}=\vev{W}_{A_{N-k}}$, that follows from the identity (\ref{conjugaterel}), $\ln \vev{W}_{A_k}$ must vanish when $k=N$, and together the identity (\ref{relsymanti}), this implies that at every order the coefficients must have a $(\pm N+k)$ factor. Similarly, $\vev{W}_{S_k}^{\text{SU}(N)}|_{N=1}=1$, so $\ln \vev{W}_{S_k}^{\text{SU}(N)}|_{N=1}=0$. Again, together with the identity (\ref{relsymanti}), this implies that at every order the coefficients must have a $(N\mp 1)$ factor, concluding the argument for (\ref{modcasimir}).

The Casimir factorization (\ref{modcasimir}) can't be true for generic representations, since the evaluation of $\ln \vev{W}_{\ydiagram{2,1}}$, eq. (\ref{perthook}), shows that it does not hold for that representation. Namely, as derived in the Appendix, $c_{\ydiagram{2,1}}=\frac{3(N^2-3)}{2N}$, but starting at order $\lambda^4$, the coefficients in the expansion of $\ln \vev{W}_{\ydiagram{2,1}}$, eq. (\ref{perthook}), are not divisible by $N^2-3$, so they don't satisfy the Casimir factorization. It would be interesting to determine if the Casimir factorization (\ref{modcasimir}) of $\ln \vev{W}_R$ holds for other representations beyond the symmetric and the antisymmetric one.

Casimir factorization bears a superficial resemblance to the hypothesis of Casimir scaling, that states that various quantities derived from vevs of logarithms of Wilson loops in QCD - chiefly the quark-antiquark static potential \cite{Ambjorn:1984dp}  - depend on the choice of representation of the matter fields only through the quadratic Casimir $c_R$. Namely, for the logarithm of the vev of a Wilson loop, 
\be
\ln \vev{W}_R  \overset{?}{=} c_R f(\lambda,N)
\ee
where $f(\lambda,N)$ is a universal function, independent of the representation $R$. In QCD, Casimir scaling of the quark-antiquark potential is known to be violated at three loops \cite{Anzai:2010td,Lee:2016cgz}. For the cusp anomalous dimension, Casimir scaling holds up to three loops \cite{Grozin:2014hna}, but in QCD is violated starting at four loops \cite{Grozin:2017css}. For $1/2$ BPS particles coupled to ${\cal N}=4$ SYM in arbitrary representations, it follows from the results of \cite{Correa:2012at} - and our expression (\ref{logofwilson}) makes it abundantly clear - that $\ln \vev{W}_R$ does not satisfy Casimir scaling, starting at four loops. Due to the relation (\ref{closedb}), it follows that the Bremsstrahlung function, and therefore the full cusp anomalous dimension, also violates Casimir scaling starting at four loops. This violation at four-loops has also been observed by explicit computation in the light-like limit of the cusp anomalous dimension \cite{Boels:2017skl}. 

It is worth emphasizing that this Casimir factorization is a property of the color invariants themselves, and in this regard, doesn't provide any information about the dynamics of the theory. On the other hand, the original Casimir scaling is a statement about the vanishing of the coefficients in front of higher order color invariants.

Finally, let's remark that this discussion was at finite $N$. In the planar limit, it follows from (\ref{altcasimir}) that for a representation $R$ whose Young diagram has $k$ boxes,
\be
c_R \rightarrow \frac{k}{2N}
\ee
so it follows from eq. (\ref{largenresult}) that 
\be
\ln \vev{W}_R^{\text{planar}} = \frac{2 c_R}{N} \ln \frac{2 I_1(\sqrt{\lambda})}{\sqrt{\lambda}}
\ee
and we conclude that in the planar limit, the ordinary Casimir scaling actually holds for $\ln \vev{W}_R$ and the quantities derived from it, like the Bremsstrahlung function $B_R$.

%	\begin{equation}
%		\langle W \rangle_0 =  1 +  \frac{g^2}{4} C_2(\mathcal{R}) + \frac{g^4}{32} %%\left(C_2(\mathcal{R})^2 - \frac{1}{6}C_2(\mathcal{R})C_2(A)   \right) + \frac{g^6}{384} \left(C_2(\mathcal{R})^3-\frac{1}{2} C_2(\mathcal{R})^2C_2(A)+ \frac{15}{12} C_2(\mathcal{R})C_2(A)^2 \right) + \cdots
%	\end{equation}
%Now it's a straightforward calculation to obtain
%	\begin{equation}
%	\log \langle W \rangle_0 \approx \frac{g^2}{4}C_2(\mathcal{R}) - \frac{g^4}{192} %C_2(A)C_2(\mathcal{R}) + \frac{g^6}{4608} C_2(A)^2 C_2(\mathcal{R}) + \cdots
%	\end{equation}
%as expected we only obtain terms proportional to $C_2(\mathcal{R})$. 

\subsection{Diagrammatic interpretation}
\begin{figure}
\centering
\includegraphics[width=.8\textwidth]{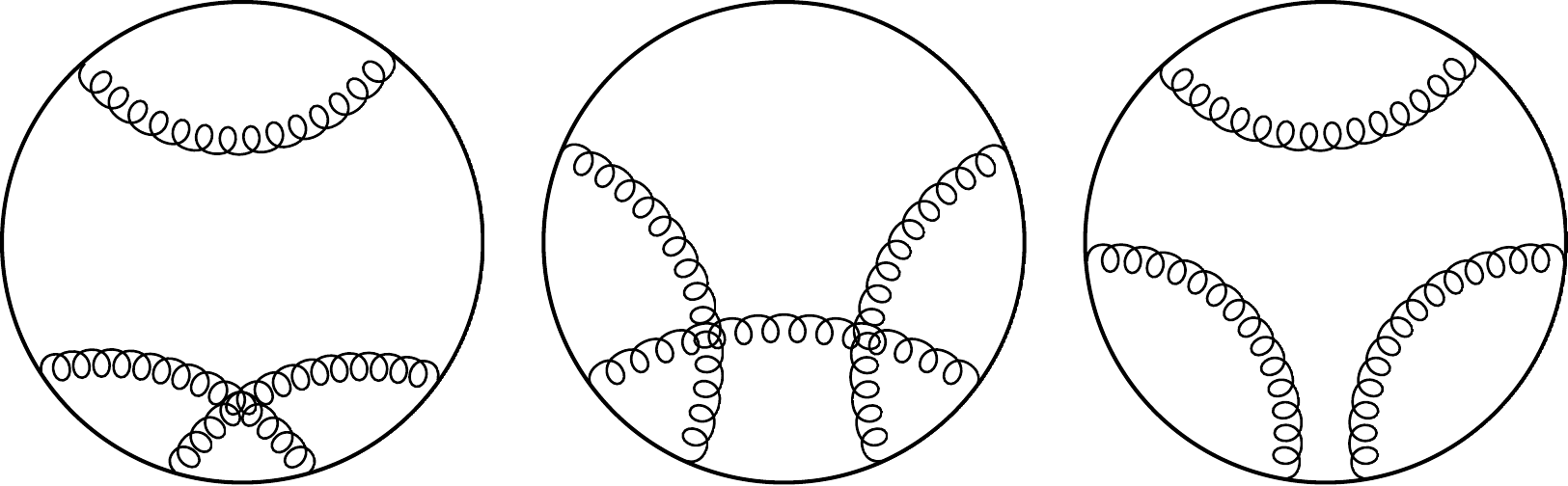}
\caption{Various examples of chord diagrams: the first one is a generic gluon diagram, they contribute to $\vev{W}_R$ at finite $N$. For $k$ gluons, by Wick's theorem there are $(2k-1)!!$ such diagrams. The second one is a connected chord diagram. They contribute to $\ln \vev{W}_R$. Their number is given by the recursion relation (\ref{connectedchord}). The last diagram is a fully disconnected chord diagram. These are the diagrams that contribute to the planar limit of $\vev{W}_{\ydiagram{1}}$ \cite{Erickson:2000af}. For $k$ gluons, there are ${\cal C}_k= \frac{(2k)!}{(k+1)! k!}$ of them.}
\label{chordzoo}
\end{figure}

We now want to provide a diagrammatic interpretation of the perturbative expansion (\ref{logofwilson}) of $\ln \vev{W}_R$. It was argued in \cite{Erickson:2000af, Drukker:2000rr} and proven in \cite{Pestun:2007rz} that in the Feynman gauge, the only Feynman diagrams that contribute to $\vev{W}_R$ involve gluon propagators starting and ending on the Wilson line. In the Mathematics literature these diagrams have been studied thoroughly, and are called chord diagrams \cite{Touchard}. At order $2n$ there are $(2n-1)!!$ of them. On the other hand, by virtue of the non-Abelian exponentiation theorem \cite{Gatheral:1983cz, Frenkel:1984pz}, to compute $\ln \vev{W}_R$ one only needs to take into account a subset of them, the so-called connected chord diagrams: diagrams where all gluon lines overlap with some other gluon line, see figure (\ref{chordzoo}).

The number of connected chord diagrams with $n$ chords satisfies the following recursion relation \cite{stein, nijenwilf}
\be
a_1=1 \hspace{1cm} a_n =(n-1)\sum_{k=1}^{n-1} a_k a_{n-k}  
\label{connectedchord}
\ee
so the first values up to the seven loops considered in this work are
\be  
a_n=1,1,4,27,248,2830,38232,\dots
\ee
It can be proven \cite{stein2} that asymptotically the ratio of the number of connected chord diagrams to the total number of chord diagrams with $n$ gluons is given by
\be 
\lim_{n\to \infty} \frac{a_n}{(2n-1)!!}=\frac{1}{e}
\ee
So, asymptotically, the number of connected Feynman diagrams is $e$ times less than the total number of Feynman diagrams. 

\begin{figure}
  \centering
  \includegraphics[width=1\textwidth]{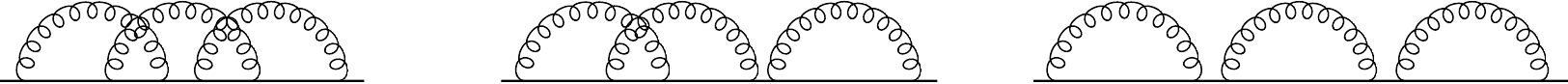}
  \put(-339,2){\Large$-$\Huge$($}
  \put(-315,2){\LARGE$2$}
  \put(-172,2){\Large$+$}
  \put(0,2){\Huge$)$}
  \put(-375,-30){\large$\bar c= c_R \left( c_R - \frac{1}{2} c_A \right)^2 -  \left( 2c_R(-\frac{1}{2}c_R c_A)+c_R^3 \right)= \frac{1}{4} c_R c_A^3$}
  \caption{Example of the determination of the modified color factor. The modified color factor of this connected diagrams with three gluons is obtained by considering the usual color factor and subtracting the color factor of all possible decompositions.}
\label{modcolorfac}  
\end{figure}

To compute $\ln \vev{W}_R$ by evaluating just the connected gluon diagrams, we have to take into account that according to the non-Abelian exponentiation theorem \cite{Gatheral:1983cz, Frenkel:1984pz}, the color factor we have to assign to each diagram is not the ordinary one, but a modified color factor $\bar c_i$. To compute $\bar c_i$ of a given connected gluon diagram, we have to consider the original color factor, and subtract the color factor of all possible decompositions of the diagram, see figure (\ref{modcolorfac}) for an illustration of this procedure.

\begin{figure}
\centering
\includegraphics[width=.8\textwidth]{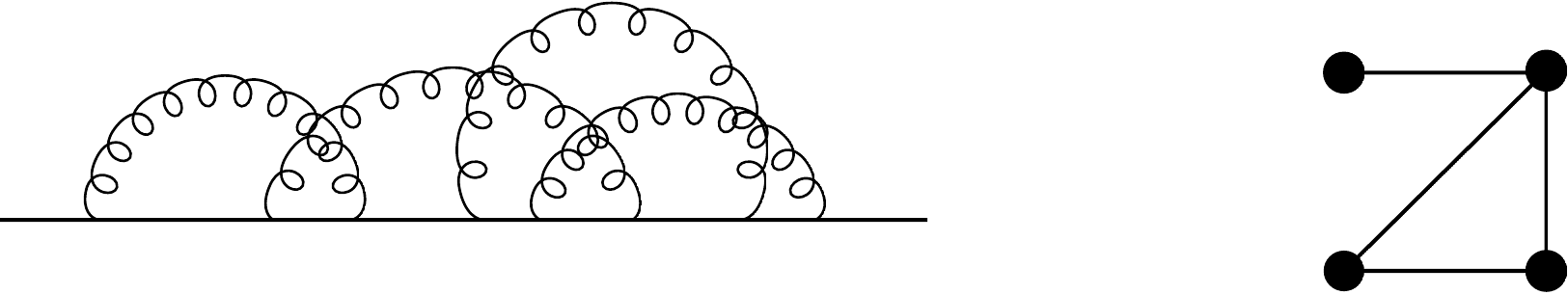}
\caption{Example of intersection graph associated to a Feynman diagram with four gluons: for each gluon, draw a dot on the plane; each time two gluon lines intersect, draw a link between the corresponding two dots.}
\label{integraph}
\end{figure}

There is a further reduction on the number of gluon diagrams that one needs to consider, since many connected chord diagrams have the same reduced color factor. The relevant object that determines whether two chord diagrams have the same reduced color factor is the intersection graph associated to a given diagram. For every chord diagram one defines an intersection graph as follows \cite{Bouchet}: for each chord introduce a point on the plane; if two chords cross, draw an edge between the two points, see figure (\ref{integraph}) for an example\footnote{Intersection graphs of chord diagrams have appeared recently in discussions of the SYK model \cite{a:2018kvh, Jia:2018ccl}.}. If the crossing graphs are isomorphic, then the reduced color factors of the original chord diagrams are the same. Since only connected chord diagrams contribute to $\ln \vev{W}_R$, we can restrict our attention to connected intersection graphs. The number of non-isomorphic connected intersection graphs for chord diagrams has been discussed in \cite{arratia}. Their numbers are
\be
1,1,2,6,21,110,789,8336,117283,\dots
\ee
So for instance, at order $g^4$, there are $7!!=105$ chord diagrams, 27 connected chord diagrams and only 6 connected intersection graphs.

\begin{figure}[ht]
  \centering
  \begin{subfigure}{0.9\linewidth}
    \includegraphics[width=.50\linewidth]{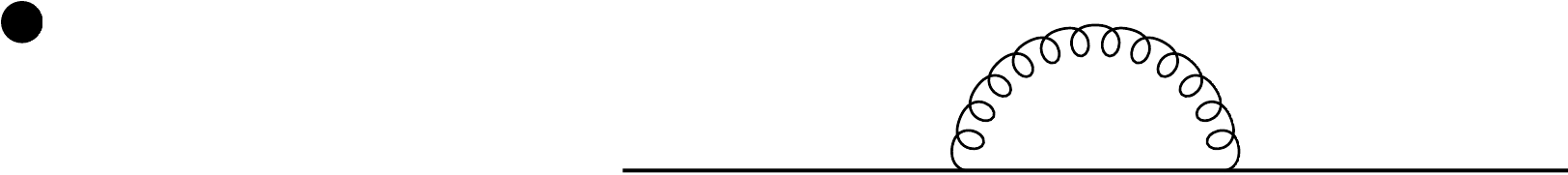}
    \put(55,15){$c_R $}
  \end{subfigure}
  \vspace*{10px}
  \begin{subfigure}{0.9\linewidth}
    \includegraphics[width=.50\linewidth]{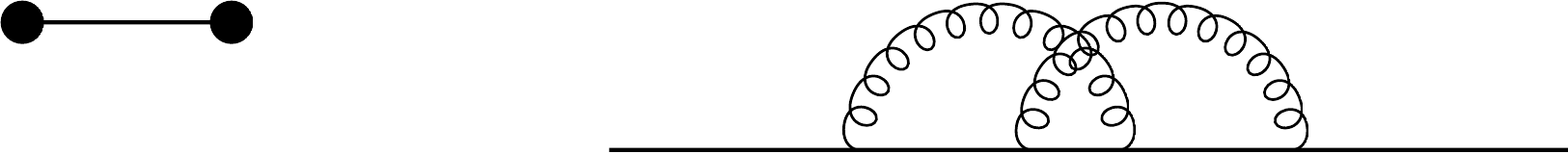}
    \put(55,15){$-\frac{1}{2}c_R c_A$}
  \end{subfigure}
  \vspace*{15px}
  \begin{subfigure}{0.9\linewidth}
    \includegraphics[width=.50\linewidth]{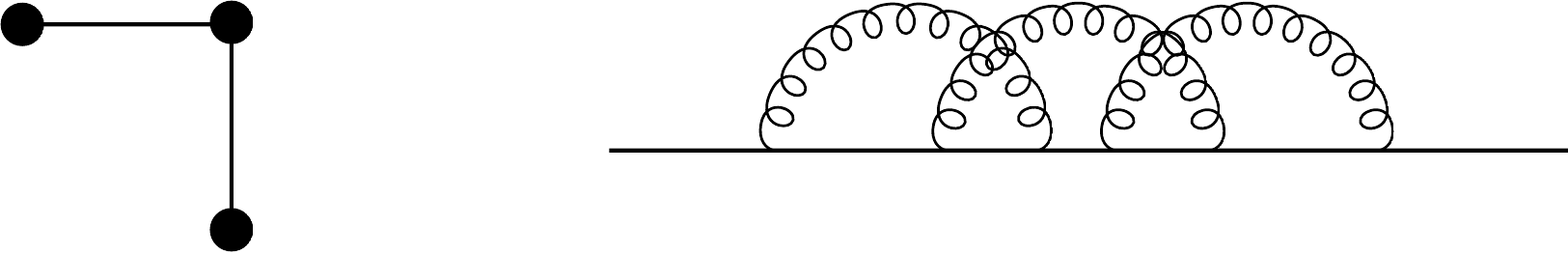}
    \put(55,15){$\frac{1}{4}c_R c_A^2$}
  \end{subfigure}
	\vspace*{10px}
  \begin{subfigure}{0.9\linewidth}
    \includegraphics[width=.50\linewidth]{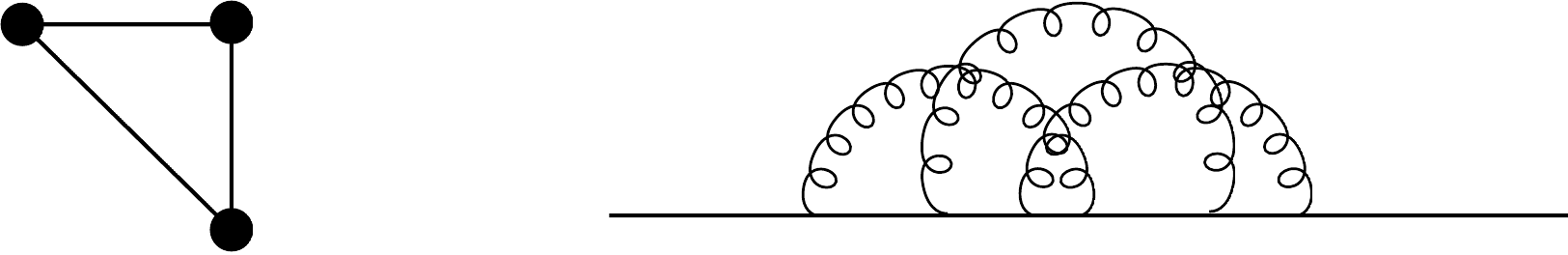}
    \put(55,15){$\frac{1}{2}c_R c_A^2$}
  \end{subfigure}  
  \vspace*{10px}
  \begin{subfigure}{.9\linewidth}
    \includegraphics[width=.50\linewidth]{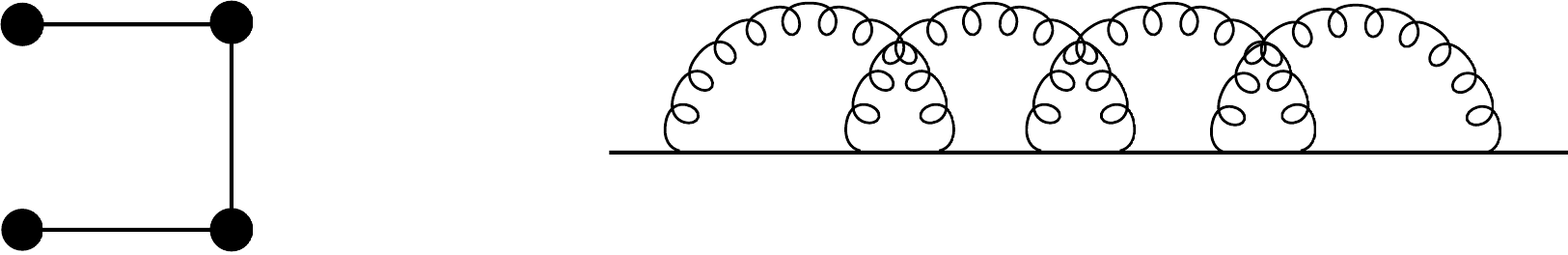}
    \put(55,15){$-\frac{1}{8}c_R c_A^3$}
  \end{subfigure}  
  \vspace*{15px}
  \begin{subfigure}{.9\linewidth}
    \includegraphics[width=.50\linewidth]{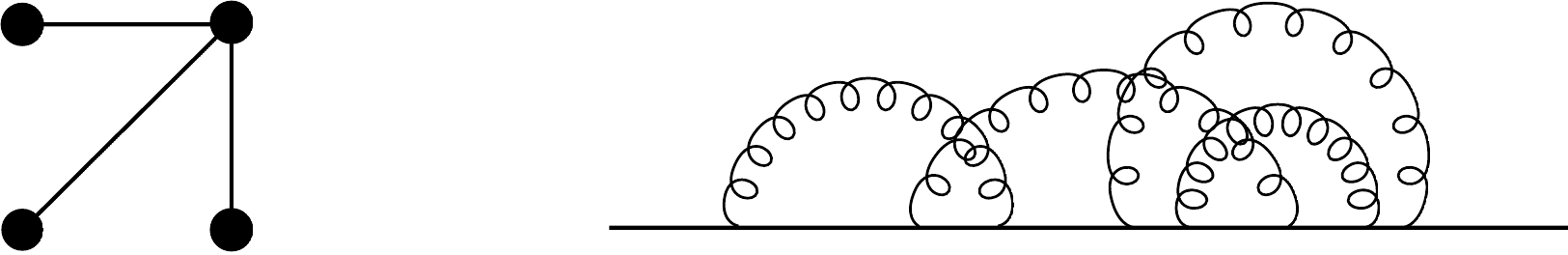}
    \put(55,15){$-\frac{1}{8}c_R c_A^3$}
  \end{subfigure}  
  \vspace*{10px}
  \begin{subfigure}{0.9\linewidth}
    \includegraphics[width=.50\linewidth]{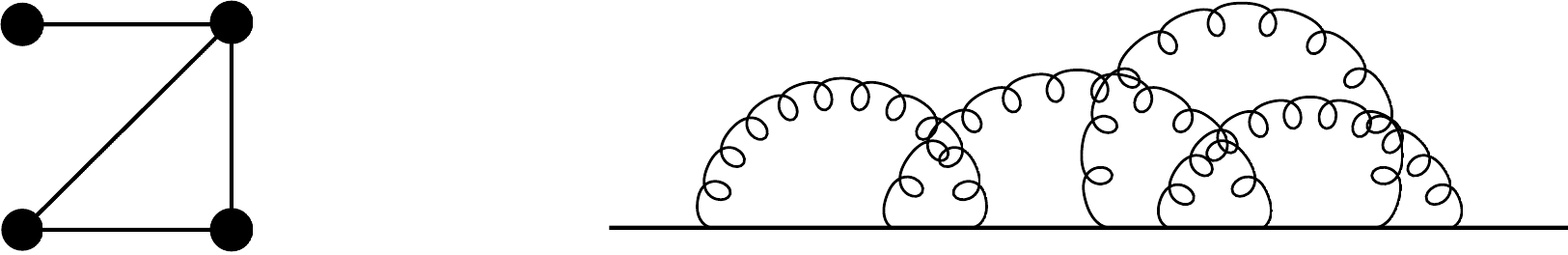}
    \put(55,15){$-\frac{1}{4}c_R c_A^3$}
  \end{subfigure}  
  \vspace*{10px}
  \begin{subfigure}{0.9\linewidth}
    \includegraphics[width=.50\linewidth]{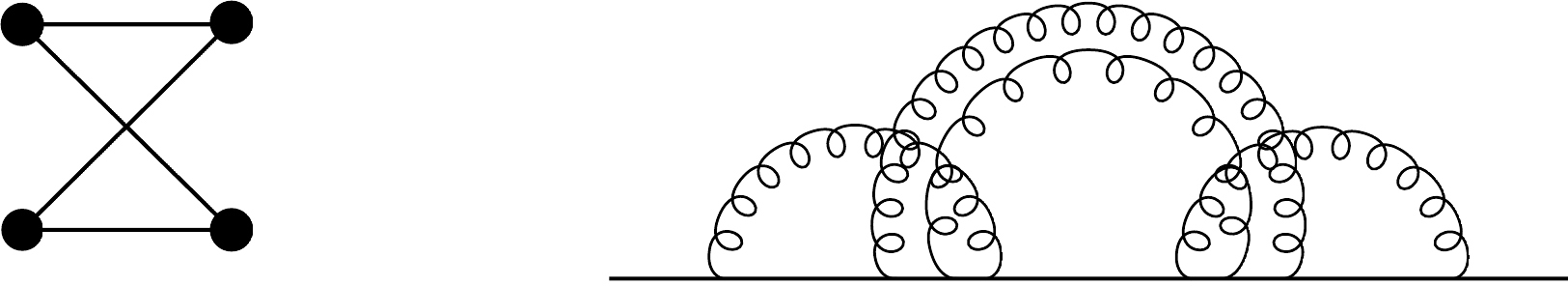}
    \put(55,15){$-\frac{5}{12}c_R c_A^3+\frac{d_R^{abcd}d_A^{abcd}}{d_R}$}
  \end{subfigure}  
  \vspace*{10px}
  \begin{subfigure}{0.9\linewidth}
    \includegraphics[width=.50\linewidth]{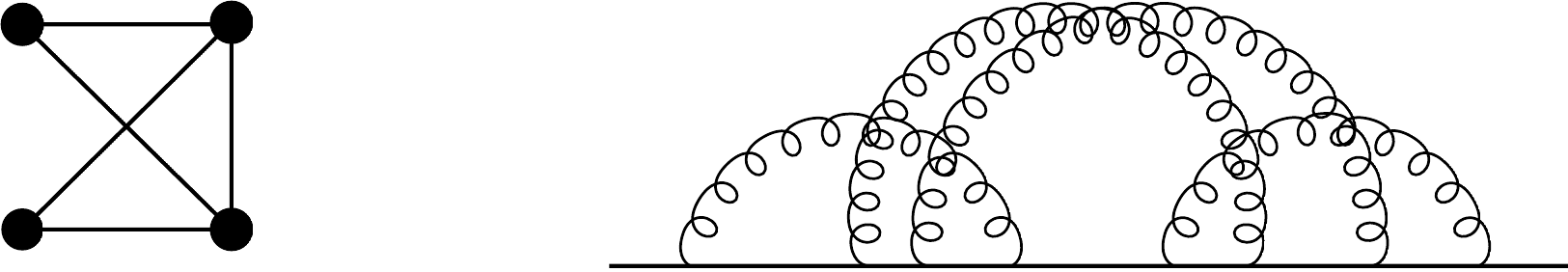}
    \put(55,15){$-\frac{13}{24}c_R c_A^3+\frac{d_R^{abcd}d_A^{abcd}}{d_R}$}
  \end{subfigure}  
  \vspace*{10px}
  \begin{subfigure}{0.9\linewidth}
    \includegraphics[width=.50\linewidth]{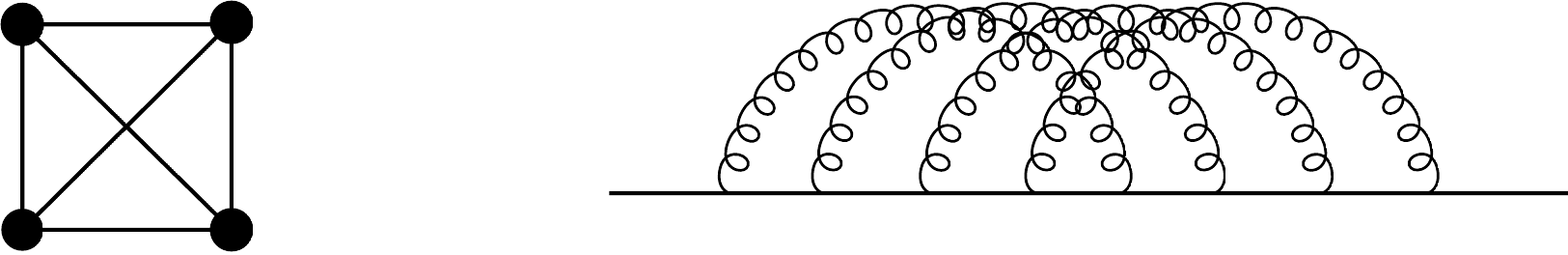}
    \put(55,15){$-\frac{19}{24}c_R c_A^3+\frac{d_R^{abcd}d_A^{abcd}}{d_R}$}
  \end{subfigure}  
  \caption{The first column displays all the non-isomorphic intersection graphs, up to four loops (graphs with four dots). The middle column shows a representative of the connected chord diagrams that share the given graph. The last column displays the modified color factor that must be assigned to all diagrams with the same intersection graph.}
\label{chordsgraphs}
\end{figure}

We currently don't know how to read off the modified color factor directly from the intersection graph. Therefore, the procedure we propose is the following: first, group all connected chord diagrams, according to their intersection graphs. For each intersection graph, evaluate the modified color factor by computing it for any of the associated connected chord diagrams. Finally, add the contributions of all connected chord diagrams. We have carried out this procedure up to four loops. The results appear in figure (\ref{chordsgraphs}). At first order there is a single diagram, with modified color factor $\bar c=c_R$. At second order there is a again a single diagram, with $\bar c=-\frac{1}{2}c_R c_A$. At third order, there are four connected chord diagrams; three of them share the first intersection graph with three dots, and have $\bar c=\frac{1}{4}c_R c_A^2$, while the fourth one has $\bar c =\frac{1}{2}c_R c_A^2$. At fourth order there are 27 connected chord diagrams, grouped according to the displayed  six intersection graphs as follows
$27=8+4+8+2+4+1$.

If $n$ is the number of gluon propagators, we have
\be
\ln \vev{W}_R= \sum_{n=1}^\infty \frac{1}{2n!} \left(\frac{{\gYM}^2}{2}\right)^n \sum_{conn} \bar c_i
\ee
where the sum $\sum_{conn}$ runs over connected chord diagrams with $n$ gluon propagators. Summing over all connected chord diagrams with up to four gluons, weighted by the modified color factors that appear in figure (\ref{chordsgraphs}), we reproduce the expansion (\ref{logofwilson}) up to fourth order.

\subsection{Comments on the coefficients}
In this concluding subsection, we indulge in a bit of numerology, and point out some patterns that we have spotted in the numerical coefficients that appear in the expansion (\ref{logofwilson}) of $\ln \vev{W}_R$. Before we proceed, we must emphasize that starting at seventh order, color invariants are not all independent; the first identity they satisfy is \cite{vanRitbergen:1998pn}
\be
d_A^{abcdef}d_A^{abcdef}-\frac{5}{8}d_A^{abcd}d_A^{cdef}d_A^{efab}+\frac{7}{240}c_A^2 d_A^{abcd}d_A^{abcd}+\frac{1}{864}c_A^6 d_A =0
\label{adjointrelation}
\ee
For this reason, starting at seventh order, one must make a choice of color invariants to present any result, and any claim about the coefficients in front of the color invariants must take this ambiguity into account.

The first observation is that, up to sixth order, the coefficients of $c_R$ in the perturbative expansion (\ref{logofwilson}) are of the form 
\be
{\cal C}_k c_R \left( \frac{-c_A}{12} \right)^k g^{k+1}
\label{catalancr}
\ee
where ${\cal C}_k$ are Catalan numbers
\be
{\cal C}_k=\frac{1}{k+1}{2k \choose k}=1,1,2,5,14,42,132,\dots
\ee
At first sight, the appearance of Catalan numbers is hardly surprising, since they are ubiquitous in combinatorial problems, and in particular in graph enumeration. In fact, Catalan numbers appear in the planar approximation of the $1-$point functions in the Hermitian Gaussian matrix model  \cite{Brezin:1977sv}
\be
\vev{tr \, \phi^{2k}}={\cal C}_k N^{k+1} +{\cal O}(N^{k-1})
\ee
or equivalently, see figure (\ref{chordzoo}), in the planar approximation to the vev of the Wilson loop in the fundamental representation, where ${\cal C}_k$ counts the number of diagrams with $k$ non-crossing gluons \cite{Erickson:2000af} 
\be
\vev{W}_{planar}=\frac{1}{N}\sum _{k=0}^\infty \vev{tr\, \phi^{2k}}_{planar} \frac{g_\text{YM}^{2k}}{(2k)!}
=\sum_{k=0}^\infty \frac{{\cal C}_k}{(2k)!}
\left( \frac{\lambda}{4}\right)^k
= \frac{2 I_1(\sqrt{\lambda})}{\sqrt{\lambda}}
\ee
However, we haven't been able to argue that the coefficients of $c_R$ in 
(\ref{logofwilson}) should follow the pattern (\ref{catalancr}). The difficulty in finding such an argument is that these coefficients arise from the interplay of combinatorics (diagram counting) and manipulations of Lie algebra generators, and we haven't managed to translate this interplay into a purely counting problem.

A second observation is that the coefficients of the $d_R^{abcd}d_A^{abcd}$ invariant, up to seventh order, follow a similar pattern, where now the numerators are given by Eulerian numbers, $A(k,1)=2^k-k-1$, 
\be 
\frac{A(k,1)}{10} c_A^{k-2}\frac{d_R^{abcd}d_A^{abcd}}{d_R} \left(\frac{-g}{6}\right)^k g^2 
\ee
A third an final observation is that, again up to sixth order, when a color invariant appears for the first time in the expansion (\ref{logofwilson}) of $\ln \vev{W}_R$, the coefficient in front of it is a unit fraction, a fraction with numerator equal to one.\footnote{To avoid confussion, the coefficients that this observation refers to include the $\frac{1}{k!}$ factor in the $g^k$ term in (\ref{logofwilson}).}

At seven loops, in the basis of color invariants chosen to present the result (\ref{logofwilson}), the pattern (\ref{catalancr}) no longer holds. However, the numerical coefficient in front of $c_R c_A^6$ comes strikingly close to follow the pattern (\ref{catalancr}), if we recall that ${\cal C}_7=132$
\be
\frac{72757}{1646023680}c_R c_A^6 = \frac{131.985\dots}{12^6}c_R c_A^6
\ee
Similarly, the third observation doesn't hold either: the color invariants that appear for the first time at seven loops in (\ref{logofwilson}) have coefficients that are not unit fractions. At this order, the second observation is not affected by the ambiguity due to the relation (\ref{adjointrelation}), but presumably at higher orders it will be affected by similar identities involving  $d_R^{a_1\dots a_k}$.  

As emphasized above, seven loops is precisely the first order where there are identities among color invariants, (\ref{adjointrelation}) being the first one. So it is natural to ask whether the breakdown of the patterns spotted up to six loops can be restored by the use of this relation. Since equation (\ref{adjointrelation}) is an identity among invariants, we can use it to impose by hand that the coefficient of $c_R c_A^6$ is indeed the one following the Catalan pattern, at the expense of introducing an overcomplete basis of color invariants. The terms that will be affected by the change are
\be 
\frac{72757}{1646023680}c_R c_A^6
+\frac{817}{16329600}c_A^2 c_R\frac{d_A^{abcd}d_A^{abcd}}{d_A}+
\frac{691}{95256000}c_R\frac{d_A^{abcd}d_A^{abef}d_A^{cdef}}{d_A}
\ee
and after the use of the identity (\ref{adjointrelation}), they turn into
\be
\frac{132}{12^6}c_Rc_A^6 + 
\frac{13}{259200}c_A^2 c_R\frac{d_A^{abcd}d_A^{abcd}}{d_A}+
\frac{1}{216000}c_R\frac{d_A^{abcd}d_A^{abef}d_A^{cdef}}{d_A}+
\frac{1}{238140}c_R\frac{d_A^{abcdef}d_A^{abcdef}}{d_A}
\ee
Notice that if we impose by hand that the pattern (\ref{catalancr}) is preserved at seventh order, it turns out that the coefficients of the color invariants that appear for the first time at this order are now unit fractions, thus restoring also the validity of the third observation at seventh order. While the relevance of this fact is unclear to us, there was no a priori reason for it to happen.

So in closing, an open question is whether at higher orders in the expansion (\ref{logofwilson}) it is always possible to use relations among color invariants to present the result in a way that the three observations presented above hold to all orders. If this turns out to be the case, a second question would be if these patterns hint at an alternative way of computing (\ref{logofwilson}), in which they are easily explained.

\acknowledgments
We would like to thank Marc Noy, Juanjo Ru\'e and Gen\'is Torrents for correspondence and explanations about chord diagrams. We would also like to thank Raimon Luna for writing a very helpful Mathematica code, and Anton Cyrol for help with FormTracer \cite{Cyrol:2016zqb}. Finally, we would like to thank the authors of FORM \cite{Vermaseren:2000nd} and FeynCalc \cite{Mertig:1990an} for making available these packages to the scientific community. Research supported by Spanish MINECO under projects MDM-2014-0369 of ICCUB (Unidad de Excelencia ``Mar\'ia de Maeztu") and FPA2017-76005-C2-P, and by AGAUR, grant 2017-SGR 754.  J. M. M. is further supported by "la Caixa" Banking Foundation (LCF/BQ/IN17/11620067), and from the European Union's Horizon 2020 research and innovation programme under the Marie Sk{\l}odowska-Curie grant agreement No. 713673. A. R. F. is further supported by an FPI-MINECO fellowship.

\appendix
\section{Color invariants}
In this Appendix we collect our conventions for color invariants, which are largely those of \cite{vanRitbergen:1998pn}. We also present the explicit results we use in the main body of the paper; some of them are already listed in \cite{Cvitanovic:2008zz, vanRitbergen:1998pn, Okubo:1983sv}. 

Let $R$ be a representation of a Lie algebra: $F$ and $A$ denote the fundamental and the adjoint representations. The dimension of $R$ is denoted by $d_R$. The generators $T_R^a$ of the representation satisfy 
\be
[T_R^a, T_R^b]=if^{abc} T_R^c \hspace{1cm} a,b =1,\dots, d_A
\ee
This does not fix the normalization of the generators $T_R^a$. We introduce two representation-dependent constants,
\begin{align*}
\hbox{tr }T_R^a T_R^b & =I_2(R) \delta^{ab} \\
T_R^aT_R^a & =c_R \mathbb{1}_{d_R\times d_R}
\end{align*}
These two representation-dependent constants are related as follows
\be
d_A I_2(R)=d_R c_R
\label{relamongcas}
\ee
In this work, we consider representations different from the fundamental only for the group SU($N$). Irreducible representations of $SU(N)$ are labelled by Young diagrams, given by $k\leq N-1$ rows of $\lambda_i$ boxes, $(\lambda_1,\dots,\lambda_k)$ with $\lambda_1\geq \lambda_2\geq \dots \geq \lambda_k$. We recall briefly how to compute the dimension and $c_R$ of a representation from its Young diagram.

One can compute the dimension of the representation $R$ of SU($N$) from the Young diagram as follows \cite{robinson}: given a box of the diagram, define its hook length $h_i$ by the number of boxes in the hook formed by the boxes to its right (in the same row), the boxes below it (in the same column), and the box itself. Then, to compute the dimension of $R$, start writing $N$ inside the box at the upper left corner of the Young diagram. Then fill the remaining boxes with numbers $N_i$, obtained by adding one every time one moves to the right, and subtracting one every time one moves down. The dimension of the representation is
\be
d_R =\prod_i \frac{N_i}{h_i}
\label{dimofrep}
\ee

\begin{figure}
\centering
\ytableausetup{boxsize=2.5em}
 \ytableausetup{textmode}
\begin{ytableau}
N & N+1 & N+2 \\
N-1 & $N$ \\
\end{ytableau}
\ytableausetup{boxsize=5pt}
\hspace{1cm}
$d_{\ydiagram{3,2}}= \frac{N(N+1)(N+2)(N-1)N}{4\times3\times1\times2\times 1}$
\caption{Example of the computation of the dimension of an irreducible representation of $SU(N)$.}
\label{exampleofnr}
\end{figure}
Figure (\ref{exampleofnr}) displays the computation of $d_R$ for a particular example. It follows from this formula that if $R$ is a representation whose Young diagram has $k$ boxes, and $R^t$ the representation with transpose Young diagram
\be
d_{R^t}(N)=(-1)^k d_R(-N)
\label{dimrtrans}
\ee
The quadratic Casimir $c_R$ for the representation with Young diagram $(\lambda_1,\lambda_2,\dots, \lambda_m)$ is given by \cite{barut}
 
\be
c_R=I_2(F) \left(\sum_{i=1}^m\lambda_i (N+\lambda_i+1-2i)-\frac{(\sum_i \lambda_i)^2}{N}\right)
\label{casofrep}
\ee
This expression can be rewritten as follows \cite{Gross:1993hu} ,
\be
c_R=I_2(F)\left(kN\, +\, \sum_i \lambda_i^2-\sum_j (\lambda_j^T)^2 \, -\frac{k^2}{N}\right)
\label{altcasimir}
\ee
where $k$ is the number of boxes of the Young diagram, $k=\sum_i \lambda_i$. This formula makes manifest that 
\be
c_{R^t}(-N)=-c_R(N)
\ee
Once one has $d_R$ and $c_R$ for a given representation R, $I_2(R)$ follows from eq. (\ref{relamongcas}).

\subsection{Higher order invariants}
Define the fully symmetrized traces as a normalized sum over all the possible index permutations
\be
d_R^{a_1\dots a_k}= \frac{1}{k!} \sum_{\sigma \in {\cal S}_k} \text{tr }\left(T_R^{a_{\sigma(1)}}\dots T_R^{a_{\sigma(k)}}\right)
\ee
It will be very useful to define the Chern character of a representation \cite{vanRitbergen:1998pn}, as a function of dummy variables $F^a$. The symmetrized traces defined above appear in the expansion of the character,

\be
ch_R (F) = tr \, e^{F^a T^a_R} =\sum_{k=0}^\infty \frac{1}{k!}d_R^{a_1\dots a_k} F^{a_1}\dots F^{a_k}
\ee

In the main body of the paper we need the evaluation of color invariants for various representations of SU($N$), and also for the fundamental representation of SO($N$). The strategy we have used is to first derive results for the fundamental representation (most of them are already available in \cite{vanRitbergen:1998pn}). For higher dimensional representations, we will first relate their Chern character to that of the fundamental representation, and then evaluate their color invariants, making use of the results found for the fundamental representation.

\subsection{Invariants for the fundamental representations of SU($N$) and SO($N$)}
The following formulas have been computed using FORM \cite{Vermaseren:2000nd}, and in the SU($N$) case, checked with FeynCalc \cite{Mertig:1990an}.

\subsubsection{Color invariants for SU($N$)}
For SU($N$) we choose the usual normalization $I_2(F)=1/2$. Then

\be
d_F=N\hspace{.5cm} c_F=\frac{N^2-1}{2N}\hspace{.5cm}
d_A=N^2-1 \hspace{.5cm}c_A=I_2(A)=N  
\ee
The relevant color invariants are
\begin{align*}
d_F^{abcd}d_A^{abcd} & =\frac{N(N^2-1)(N^2+6)}{48} \\
d_A^{abcd}d_A^{abcd} & =\frac{N^2 (N^2-1)(N^2+36)}{24} \\
%\be
%d_F^{abcd}d_F^{cdef}d_F^{efab}=\frac{(N^2-1)(N^2-3)(N^4-6N^2+63)}{(12N)^3}
%\ee
%\be
%d_F^{abcd}d_F^{cdef}d_A^{efab}=\frac{(N^2-1)(N^4+12N^2-9)}{864}
%\ee
d_F^{abcd}d_A^{cdef}d_A^{efab} & = \frac{N^3(N^2-1)(N^2+51)}{432} \\
d_A^{abcd}d_A^{cdef}d_A^{efab} & =\frac{N^2(N^2-1)(N^4+135N^2+324)}{216} \\
d_F^{abcdef}d_A^{abcdef} & =\frac{(N^2-1) N (N^4+36N^2+120)}{3840} \\
d_F^{abcdef}d_A^{abcg}d_A^{defg} & =\frac{N^2(N^2-1)(N^4+45N^2+84)}{3840} \\
d_A^{abcdef}d_F^{abcg}d_A^{defg} & =\frac{N^2(N^2-1)(N^4+141N^2+540)}{3840}
\end{align*}
%\be
%d_F^{abcdef}d_F^{abcg}d_A^{defg}=-\frac{(N^2-1)N(N^4+57N^2+36)}{7680}
%\ee

\subsubsection{Color invariants for SO($N$)}
For SO($N$) we choose the usual normalization $I_2(F)=1$. Then
\be 
d_F=N \hspace{.5cm} c_F=\frac{N-1}{2} \hspace{.5cm}
N_A=\frac{N(N-1)}{2}\hspace{.5cm}c_A=N-2
\ee
The relevant color invariants are
\begin{align*}
 d_F^{abcd}d_A^{abcd} & =\frac{d_A c_A}{24}(N^2-7N+22) \\
 d_A^{abcd}d_A^{abcd} & =\frac{d_A c_A}{24} (N^3-15N^2+138N-296) \\
%\be
%d_F^{abcd}d_F^{cdef}d_F^{efab}=\frac{N(N-1)(2N^3-3N^2+33N-16)}{864}
%\ee
%\be
%d_F^{abcd}d_F^{cdef}d_A^{efab}=\frac{N(N-1)(N-2)(2N^3-15N^2+111N-142)}{864}
%\ee
 d_F^{abcd}d_A^{cdef}d_A^{efab} & =\frac{d_A c_A}{432}(2N^4-31N^3+387N^2-1582N+2048) \\
 d_A^{abcd}d_A^{cdef}d_A^{efab} & =\frac{d_A c_A}{432}(2N^5-47N^4+971N^3-7018N^2+23272N-29440) \\
 d_F^{abcdef}d_A^{abcdef} & =\frac{d_A c_A}{960} \left(N^4-32N^3+273N^2-902N+1312\right) \\
 d_F^{abcdef}d_A^{abcg}d_A^{defg} & =\frac{d_A c_A}{960}\left(
N^5-16N^4+193N^3-1214N^2+3656N-3920\right) \\
 d_A^{abcdef}d_F^{abcg}d_A^{defg} & =\frac{d_A c_A}{960}\left(
N^5-40N^4+697N^3-4598N^2+14576N-17888\right)
\end{align*}

\subsection{Invariants for $S_k/A_k$ representations of SU($N$)}
Applying the formulas (\ref{dimofrep}) and (\ref{casofrep}) for the representations $A_k/S_k$ of SU($N$), we have
\begin{align*}
d_{{\cal A}_k} & ={N \choose k}   & c_2({\cal A}_k)   =I_2(F) \frac{k(N+1)(N-k)}{N}  & \hspace{.5cm} I_2({\cal A}_k)  =I_2(F){N-2 \choose k-1} \\
d_{S_k} & ={N+k-1 \choose k}   & c_2(S_k)  =I_2(F) \frac{k(N-1)(N+k)}{N} 
  & \hspace{.5cm} I_2(S_k)  =I_2(F){N+k \choose k-1}
\end{align*}

We now turn to color invariants involving higher order symmetrized traces $d_R^{a_1\dots a_m}$, for $R=S_k/A_k$. For invariants involving up to $d_R^{abcd}$, we first derive the formulas valid for arbitrary $k$, and then check them via an alternative computation, for $k=1,2,3,4$.

For invariants involving $d_R^{abcdef}$, we have explicitly computed the results for $k=1,2,3,4$, and then we have guessed a formula for generic $k$, imposing that the formulas are invariant under $k\rightarrow N-k$ for $A_k$. So the formulas quoted have been only derived for $k=1,2,3,4$ but are probably true also for any $k$. According to \cite{Okubo:1983sv}
\be
d_{S_k}^{abcd}=\frac{N(N-1)+6k (N+k)}{(k-1)!(N+3)!}(N+k)! d_F^{abcd}+
{N+k+1 \choose k-2}I_2(F)^2\left(\delta^{ab}\delta^{cd}+\delta^{ac}\delta^{bd}+\delta^{ad}\delta^{bc}\right)
\label{dsk4}
\ee

\be
d_{A_k}^{abcd}=\frac{N(N+1)-6k (N-k)}{(k-1)!(N-k-1)!}(N-4)! d_F^{abcd}+
{N-4 \choose k-2}I_2(F)^2\left(\delta^{ab}\delta^{cd}+\delta^{ac}\delta^{bd}+\delta^{ad}\delta^{bc}\right)
\label{ask4}
\ee
To obtain the relevant color invariants, we contract these formulas with various symmetrized traces in the adjoint representation, and use
\be
d_A^{aacd}=\frac{5}{6}c_A^2 \delta^{cd}
\ee
\be
d_A^{aacdef}=\frac{7}{10}c_Ad_A^{cdef}
\ee
The results are as follows (the upper sign is for $S_k$, the lower one is for $A_k$), 
\begin{align*}
 d_R^{abcd} d_A^{abcd} &  = c_R d_R\frac{N}{24}\left(N^2\mp 6N +6k(k\pm N)\right) \\
 d_R^{abcd} d_A^{abef} d_A^{cdef} &  = c_R d_R \frac{N^2}{216} \left(N^3\mp 6N^2 -9N\mp 54 +6Nk(k\pm N) +54 k (N\pm k)\right)  \\
  d_A^{abcdef}d_R^{abcg}d_A^{defg} & = c_R d_R \frac{N^2}{1920}\left(N^4\mp 6N^3+81N^2\mp 594N +(N^2\pm 54N+540)k(k\pm N)\right)  \\
\end{align*}
Note that they satisfy the $N\rightarrow -N$ symmetry when $S_k\rightarrow A_k$ (up to global sign) and for $A_k$ the $k\rightarrow N-k$ symmetry. In order to repeat the same procedure to evaluate color invariants involving $d_{S_k/A_k}^{abcdef}$, we would need formulas similar to eqs. (\ref{dsk4}) and (\ref{ask4}) for $d_{S_k/A_k}^{abcdef}$. From \cite{Okubo:1983sv} one can derive the leading terms in such formulas
\begin{align*}
d_{S_k}^{abcdef} & = \sum_{i=0}^{k-1}(k-i)^5 {N+i-1 \choose i}d_F^{abcdef} + \dots \\
d_{A_k}^{abcdef} & = \sum_{i=0}^{k-1} (-1)^{k-1-i} (k-i)^5 {N \choose i}d_F^{abcdef} + \dots
\end{align*}
but we are not aware of complete formulas for arbitrary $k$. Instead, we will compute them for small values of $k$, from the character formulas for the symmetric and antisymmetric representations 
\cite{vanRitbergen:1998pn}
\begin{align*}
Ch_{S_k}(F) & =\sum_{\substack{n_i,m_i \\ k=n_i m_i}} 
\prod_i \frac{1}{m_i!}\left(\frac{Ch(n_iF)}{n_i}\right)^{m_i} \\
Ch_{A_k}(F) & =(-1)^k \sum_{\substack{n_i,m_i \\ k=n_i m_i}} 
\prod_i \frac{1}{m_i!}\left(-\frac{Ch(n_iF)}{n_i}\right)^{m_i}
\end{align*}
where the sum is over all partitions of $k$ into different integers $n_i$, each appearing with multiplicity $m_i$. From these formulas, we obtain the characters of $S_k,A_k$ for $k=2,3,4$,
\begin{align*} 
Ch_{S_2/{\cal A}_2}(F) & =\frac{1}{2}(Ch\, F)^2\pm \frac{1}{2}Ch(\, 2F) \\
Ch_{S_3/A_3}(F) & =\frac{1}{3!}(Ch\, F)^3\pm\frac{1}{2}Ch\, 2F Ch\, F +\frac{1}{3} Ch\, 3F \\
Ch_{S_4/A_4}(F) & =\frac{1}{4!}(Ch\, F)^4\pm\frac{1}{4}Ch\, 2F (Ch\, F)^2 +\frac{1}{8}(Ch\, 2F)^2+\frac{1}{3} Ch\, 3F Ch\, F \pm \frac{1}{4} Ch\, 4F
\end{align*}

We expand in powers of $F$ up to sixth order. At zeroth, second and fourth orders we recover the formulas for $N_{S_k/A_k}$, $c_{S_k/A_k}$ and $d_{S_k/A_k}^{abcd}$ for $k=2,3,4$. At sixth order, we obtain the following formulas for $d_{S_k/A_k}^{abcdef}$,

\be 
d_{S_2/{\cal A}_2}^{abcdef} =
(N\pm 32)d_F^{abcdef}+I_2(F)\left(\delta^{ab}d_F^{cdef}+\dots\right)+
\left(d_F^{abc}d_F^{def}+\dots\right)
\ee

\begin{multline} 
d_{S_3/A_3}^{abcdef}=\frac{N^2\pm 65 N+486}{2}d_F^{abcdef}+(N\pm 10)I_2(F)
(\delta^{ab}d_F^{cdef}+\dots)+(N\pm 8)(d_F^{abc}d_F^{def}+\dots) \\
+I_2(F)^3 (\delta^{ab}\delta^{cd}\delta^{ef}+\dots)
\end{multline}
and
\begin{multline}
d_{S_4/A_4}^{abcdef}= \frac{N^3 \pm 99 N^2+ 1556N\pm 6144}{6}d^{abcdef}_F+\frac{N^2\pm 21 N+92}{2}I_2(F)(\delta^{ab}d^{cdef}+\dots)+\\
\frac{N^2\pm 17 N+68}{2}(d^{abc}_Fd^{def}_F+\dots)+(N\pm 6)I_2(F)^3(\delta^{ab}\delta^{cd}\delta^{ef}+\dots)
\end{multline}
Using these expressions, we evaluate the following color invariants,
\begin{multline}
d_{S_k/A_k}^{abcdef}d_A^{abcdef} =  \frac{c_R d_R N}{1920} 
\left( N^4\mp 30N^3 +186 N^2 \right. \\
\left.
\pm \, 60N+30N^2k(k\pm N) +120k^2(N\pm k)^2 -300 N k (N\pm k)
\right)
\end{multline}
\begin{multline}
d_{S_k/A_k}^{abcdef}d_A^{abcg}d_A^{defg}= \frac{c_R d_R N^2}{1920} 
\left( N^4 \mp 30N^3 + 105N^2\pm 150 N + 144 \,\, \, + \right. \\
\left.  
30N^2 k(k\pm N)+120k^2 (N\pm k)^2 -210 Nk(N\pm k)-180k(k\pm N)
\right)
\end{multline}

\ytableausetup{boxsize=5pt}

We emphasize that these last two formulas have been proven only for $k=1,2,3,4$, although we are confident that they are true for arbitrary $k$. We find that all the color invariants we have computed for $S_k$ and $A_k$ are related by $N \rightarrow -N$, as expected \cite{Cvitanovic:1982bq}. 

We can perform some checks for specific values of N. For SU(4) the invariants for $A_2$ coincide with those of SO(6) in the fundamental.

\ytableausetup{boxsize=6pt}
\subsection{Results for the $\ydiagram{2,1}$ representation of SU($N$)}
In the main body of the paper, we display various results for the $\ydiagram{2,1}$ representation, since it is the simplest representation that is not fully symmetric or fully antisymmetric. Furthermore, its Young diagram is self-transpose, thus it allows to illustrate the $1/N^2$ expansion of $\vev{W}_R$ for these representations. Some of the results we need are already available in \cite{Okubo:1983sv}, but we have derived all the formulas below independently and checked them with \cite{Okubo:1983sv} when possible.

\ytableausetup{boxsize=5pt}
To obtain the character for this representation, we recall
\be
\ydiagram{1}\times \ydiagram{1}\times \ydiagram{1} =
\ydiagram{3}+2 \, \, \ydiagram{2,1}+\ydiagram{1,1,1}
\ee
from which we deduce
\be
Ch_{\ydiagram{2,1}} F =\frac{1}{3}(Ch\, F)^3-\frac{1}{3}Ch\, 3F
\ee
Expanding this result up to sixth order in $F$ we obtain
\be
d_{\ydiagram{2,1}}=\frac{N(N^2-1)}{3}\hspace{.5cm}
c_{\ydiagram{2,1}}=\frac{3(N^2-3)}{2N}\hspace{.5cm}
I_2(\ydiagram{2,1})=\frac{N^2-3}{2}
\ee

\be
d_{\ydiagram{2,1}}^{abcd}=(N^2-27)d_F^{abcd} +2N I_2(F)^2\left(\delta^{ab}\delta^{cd}+\delta^{ac}\delta^{bd}+
\delta^{ad}\delta^{bc}\right)
\ee

\be
d_{\ydiagram{2,1}}^{abcdef}=(N^2-3^5)d_F^{abcdef}+2N(d^{ab}d^{cdef}+\dots)+2N(d^{abc}d^{def}+\dots)+2(d^{ab}d^{cd}d^{ef}+\dots)
\ee
The results for $d_{\ydiagram{2,1}}$ and $c_{\ydiagram{2,1}}$ can also be derived from the general formulas (\ref{dimofrep}) and (\ref{casofrep}). With these formulas we derive the following color invariants
\begin{align*}
d^{abcd}_{\ydiagram{2,1}}d^{abcd}_A & = \frac{N (N^2-1) (N^4+39N^2-162)}{48} \\
d^{abcd}_{\ydiagram{2,1}}d^{abef}_A d^{cdef}_A & = \frac{N^3 (N^2-1)(N^4+192N^2-729)}{432} \\
d^{abcdef}_{\ydiagram{2,1}}d^{abcdef}_A & =\frac{N(N^2-1)(N^6+213N^4+6492N^2-29160)}{3840} \\
d^{abcdef}_{\ydiagram{2,1}}d^{abcg}_A d^{defg}_A & = \frac{N^2 (N^2-1)(N^6+402N^4+1389N^2-11772)}{3840} \\
d^{abcdef}_A d^{abcg}_A d^{defg}_{\ydiagram{2,1}} & = \frac{N^2 (N^2-1)(N^6+282N^4+2781N^2-14580)}{3840}
\end{align*}
We can provide two checks for these results. First, all the color invariants have a $1/N^2$ expansion, as expected since the Young diagram $\ydiagram{2,1}$ is self-transpose. Also, for SU(2), the invariants evaluate to the same numbers if we replace $d_{\ydiagram{2,1}}^{abcd},d_{\ydiagram{2,1}}^{abcdef}$ by $d_F^{abcd},d_F^{abcdef}$. 

\bibliographystyle{JHEP}

\end{document}